\documentclass{article}
\usepackage{amssymb}

\usepackage{amsmath}

\begin{document}

\begin{flushright}
\begin{minipage}[t]{55mm}
Preprint of P.N. Lebedev Physical Institute, No. 27, 1974, unpublished
\end{minipage}
\end{flushright}
\vspace{0.5cm}
\begin{center}
\LARGE{Electron Scattering by a Solenoid}\\
\vspace{0.5 cm}
\large{I.V. Tyutin\footnote{E-mail: tyutin@lpi.ru}}
\end{center}

\begin{abstract}
The quantum-mechanical problems of electron scattering by an infinitely thin
solenoid and by a half of an infinitely thin solenoid are examined from the
viewpoint of constructing a self-adjoint Hamiltonian. It is demonstrated
that in both problems there exist unique self-adjoint operators with a
``non-singular'' domain, that, due to physical reasons, are identified with
the corresponding Hamiltonians. In the case of quantized values of magnetic
flow along the solenoid, the electron does not experience any scattering by
the string. It is shown that the scattering amplitude and wave function of
an electron in the problem of scattering by an infinitely long solenoid of
radius $a$ in the limit $a\rightarrow 0$ turn into the corresponding
expressions for the problem of an infinitely thin solenoid. In particular,
at a quantized value of magnetic flow along the solenoid, scattering
disappears at $a\rightarrow 0$.
\end{abstract}

\section{Introduction}

In this article, we examine two quantum-mechanical problems:\ the scattering
of an electron by the field of an infinite, infinitely thin solenoid, $%
A_{k}^{\left( c\right) }\left( \vec{r}\right) =-\mu \frac{\varepsilon
_{3ki}r_{i}}{\rho ^{2}}$, $\vec{r}=\left( x,y,z\right) $, $\rho
^{2}=x^{2}+y^{2}$, and the scattering of an electron by the field of a
semi-infinite, infinitely thin solenoid, $A_{k}^{\left( D\right) }\left(
\vec{r}\right) =-\mu \frac{\varepsilon _{3ki}r_{i}}{r\left( r-z\right) }$.
Our interest in these problems is due to the following reasons.

As is well-known, Dirac \cite{1} proposed a Lagrangian theory of
electrodynamics with two charges:\ the electric and magnetic ones, subject
to a charge quantization condition; see \cite{1}. In Dirac's theory,
however, the electromagnetic potentials contain singularities on certain
lines (Dirac's strings) coming out of the magnetic charges (in principle,
Dirac's strings may also come out of the electric charges, as well as out of
the charges of both types). The motion of these strings is not determined by
the Lagrangian and may be defined arbitrarily. Dirac presented some
arguments in favour of the fact that physical results must actually be
independent of the form and specific motion of strings, since the
electromagnetic potentials for two different positions of a string are
related by a gradient transformation (anywhere outside strings). This
argument is not very convincing, because, in fact, two potentials for two
different positions of strings cannot be related by a gradient
transformation in the entire space (for more details, see \cite{1}).

The simplest problem concerning the interaction of an electron with a
magnetic charge is the problem of the scattering of a non-relativistic
electron by the static field of an infinitely heavy magnetic charge. In
addition, the electromagnetic potential (suggested by Dirac) of a magnetic
charge is identical with the electromagnetic potential of a semi-infinite,
infinitely thin solenoid. According to Dirac's hypothesis, physical results
(in particular, the cross-section of electron scattering) must not depend on
the direction of a string (the above expression for $A_{k}^{\left( D\right)
} $ corresponds to the rectilineal form of a string; of course, physical
results must not depend on the form of a string, either). Such a problem has
been examined by numerous authors \cite{1}. Finally, in the work by
Zwanziger \cite{2} it was found that the cross-section of electron
scattering by the potential of a magnetic charge is, in fact, independent of
the direction of a string.

If the electron ``does not feel'' the magnetic field along the string, then
it must not experience any scattering by the potential of an infinitely thin
and infinitely long solenoid (in case the magnetic flow along the string
takes quantized values, which is in agreement with the quantization
condition for the electric and magnetic charges). This problem has been
examined (in a different connection) by Aharonov and Bohm \cite{3}, who have
found that in the case of quantized values of the magnetic charge of a
solenoid (this corresponds to $\mu =0,\pm 1,\pm 2,\ldots $ in the above
expression for $A_{k}^{\left( c\right) }$), the electron, indeed, does not
experience any scattering by the string.

If, however, one looks at the structure of solutions of this problem, it
turns out that the wave function of an electron has the form $\psi =e^{i\mu
\varphi }e^{i\vec{k}\vec{r}}$, where $\vec{k}$ is the momentum of a moving
electron, and $\varphi $ is the angle between the projections of the vectors
$\vec{k}$ and $\vec{r}$ on the plane $xy$. This form of a wave function
suggests the idea that the potential $A_{k}^{\left( c\right) }$ is a pure
gradient: $A_{k}^{\left( c\right) }\thicksim \mu \partial _{k}\varphi $. If,
indeed, one formally (i.e., without taking into account the fact that $%
\varphi $ is a discontinuous function) calculates the gradient of $\varphi $%
, then one obtains the equality $A_{k}^{\left( c\right) }=\mu \partial
_{k}\varphi $. It is clear, nonetheless, that such an equality cannot take
place, since the $\mathrm{rot}$ of its r.h.s. is equal to zero, while at the
same time $\mathrm{rot\,}A_{k}^{\left( c\right) }$ is the magnetic field
(with a finite flow $\mu $)\ along the string. Then, there arises the
question as to the correctness of the solution presented by Aharonov and
Bohm for this problem.

In the case of electron scattering by a semi-infinite, infinitely thin
solenoid, Zwanziger \cite{2} has found that the scattering amplitude has the
form $f\left( \vec{k},\vec{k}^{\prime }\right) =e^{i\Omega }f_{0}\left( \vec{%
k},\vec{k}^{\prime }\right) $ ($\vec{k}$ and $\vec{k}^{\prime }$ are the
respective initial and final momenta of an electron; $f_{0}$ is the
scattering amplitude in case the string is directed along the momentum of a
moving electron) and depends, as a consequence, merely on the angle between $%
\vec{k}$ and $\vec{k}^{\prime }$; $\Omega $ is a certain function (see
Section 4). As will be shown in Section 4, the wave function $\psi
=e^{i\Omega }\psi _{0}$ has the same form, where $\psi _{0}$ is the wave
function in case the string is directed along $\vec{k}$. Once again, there
arises the suspicion that such a form of solution is due to the fact that
the difference of vector potentials $A_{i}^{\left( D\right) }-A_{i}^{\left(
D,k\right) }$ ($A_{i}^{\left( D,k\right) }\left( \vec{r}\right) =-\mu \frac{%
\varepsilon _{ilj}k_{l}r_{j}}{r\left( kr-\vec{k}\vec{r}\right) }$ is the
potential of a semi-infinite, infinitely thin solenoid, with the string
being directed along $\vec{k}$) is a pure gradient. If one formally computes
the gradient of $\Omega $ (i.e., without taking into account the fact that $%
\Omega $ is a discontinuous function), then one obtains $A_{i}^{\left(
D\right) }-A_{i}^{\left( D,k\right) }=\partial _{i}\Omega $. However, this
equality cannot take place, either, because the $\mathrm{rot}$ of the r.h.s.
is equal to zero, while at the same time the $\mathrm{rot}$ of the l.h.s. is
the flow of magnetic field (with finite flows)\ along the strings. Thus, in
this case there also arises the question as to the correctness of the
solution.

The question of correctness of the solutions found by Aharonov--Bohm and
Zwanziger for the problems in question arises also in connection with
another reason. The point is that the Hamiltonians of these problems are
extremely singular and thus cannot be defined immediately in the class of
functions for which they would be self-adjoint operators (in particular, the
natural domain of a Hamiltonian does not include all differentiable
functions). In the natural domain, the Hamiltonians prove to be merely
symmetric, with non-vanishing deficiency indices. In addition, it is
well-known \cite{4,5} that the problem of constructing a self-adjoint
Hamiltonian from a given symmetric operator (a self-adjoint extension of a
symmetric operator) admits more than one solution. Let us remind that the
self-adjoint character of an operator is necessary for the corresponding
operator of evolution to be unitary and uniquely defined in the entire
Hilbert space. Different self-adjoint extensions lead to different solutions
of the scattering problem, corresponding to the same Schr\"{o}dinger
equation.

In Sections 2 and 4, in the cases of infinite and semi-infinite solenoids
respectively, with arbitrary values of $\mu $, it is shown that in both
problems there exist unique self-adjoint extensions of the Hamiltonian, such
that they obey the usual physical condition:\ the Hilbert-space functions
for which the Hamiltonian is defined must not be singular (by the way, it is
natural to call this condition the ``principle of minimal singularity'').
Given this, it is shown that for the integer values of the parameter $\mu $
any scattering by the string is absent and the solution of the scattering
problem is identical with the corresponding solutions of Aharonov--Bohm and
Zwanziger. Therefore, in both problems it is rigorously proved that in the
case of a quantized magnetic flow along the strings the electron does not
experience scattering, in agreement with Dirac's hypothesis, and that, in
consequence, a semi-infinite, infinitely thin solenoid can actually be
considered as a realization of a fixed monopole (magnetic charge). In the
same sections, it is shown how to solve the (seeming) difficulty that has
been described earlier in this Introduction, i.e., the one related with the
gauge transformations of potentials and with the form of solutions.

In Section 3, it is shown that the result of Aharonov--Bohm can be deduced
with the help of a ``physical'' regularization of the potential, which
consists in using a solenoid with a finite radius $a$, rather than an
infinitely thin solenoid, and then making $a$ go to zero.

In Appendix A, in connection with the article \cite{7}, it is shown that the
problem of scattering by the potential of an infinitely thin solenoid cannot
be solved as a perturbation theory in $\Delta \mu $, where $\mu =[\mu
]+\Delta \mu $, and $[\mu ]$ is the integer part of $\mu $.

\section{Electron scattering by the field of an infinitely thin and
infinitely long solenoid}

The electromagnetic potential $A_{k}^{(c)}$ of an infinitely thin and
infinitely long solenoid has the form%
\begin{equation}
A_{k}^{(c)}\left( \vec{r}\right) =-\mu \frac{\varepsilon _{3ki}r_{i}}{\rho
^{2}}\,.  \label{1}
\end{equation}

It is easy to see that outside the $z$ axis the magnetic field $\vec{H}=%
\mathrm{rot\,}\vec{A}^{(c)}$ is equal to zero everywhere. Nevertheless, the
flow $\oint \vec{A}^{(c)}d\vec{l}$ through any surface with the boundary
being any contour (even an infinitely small one) around the $z$ axis does
not vanish and is equal to $\mu $.

The Schr\"{o}dinger equation for the scattering of an electron by this same
potential has the form%
\begin{equation}
i\partial _{t}\psi \left( \vec{r},t\right) =H\psi \left( \vec{r},t\right) \,,
\label{2}
\end{equation}%
where%
\begin{equation}
H=\left( -i\partial _{k}+A_{k}^{(c)}\left( \vec{r}\right) \right) ^{2}\,,
\label{3}
\end{equation}%
and the mass of an electron is set equal to $1/2$.

Eq. (\ref{2}) has the following solution:%
\begin{equation}
\psi \left( \vec{r},t\right) =\exp \left\{ -ik_{z}^{2}t+ik_{z}z\right\} \psi
\left( \vec{\rho},t\right) \,,  \label{4vopros}
\end{equation}%
where $k_{z}\,$is the projection of the momentum of the electron on the $z\,$%
axis, $\vec{\rho}=\left( x,y\right) $, and $\psi \left( \vec{\rho},t\right) $
obeys the equation%
\begin{eqnarray}
&&\,i\partial _{t}\psi \left( \vec{\rho},t\right) =\widetilde{H}\psi \left(
\vec{\rho},t\right) \,,  \label{2prime} \\
&&\,\widetilde{H}=-\Delta +\frac{2\mu i}{\rho ^{2}}\left( y\partial
_{x}-x\partial _{y}\right) +\frac{\mu ^{2}}{\rho ^{2}}  \notag \\
&&\,=-\partial _{\rho }^{2}-\frac{1}{\rho }\partial _{\rho }+\frac{1}{\rho
^{2}}\left( -\partial _{\varphi }^{2}-2i\mu \partial _{\varphi }+\mu
^{2}\right) \,.  \label{3prime}
\end{eqnarray}

Let us now construct a self-adjoint Hamiltonian corresponding to the
differential expression (\ref{3prime}).

It is clear that $\widetilde{H}$ cannot be determined immediately for any
differentiable (and decreasing sufficiently fast as $\rho \rightarrow \infty
$) functions that belong to $L_{2}$ ($L_{2}$ is the Hilbert space of
square-integrable functions $\psi \left( \vec{\rho}\right) $) because of the
terms $\frac{\mu ^{2}}{\rho ^{2}}$ and $\frac{1}{\rho ^{2}}\left( y\partial
_{x}-x\partial _{y}\right) $.

Let us define an operator $H_{0}$ as a differential operator (\ref{3prime})
on the domain $D_{0}$, where $D_{0}$ is the set of twice-differentiable
functions with a compact support that turn to zero at $\rho <a$ for a
certain $a$ ($a$ may be different for different functions). $D_{0}$ is dense
in $L_{2}$; (\ref{3prime}) is obviously defined for any $\psi \in D_{0}$,
and for any $\psi _{1},\psi _{2}\subset D_{0}$ there holds the equality%
\begin{equation}
\int d\vec{\rho}\psi _{1}^{\ast }\left( H_{0}\psi _{2}\right) =\int d\vec{%
\rho}\left( H_{0}\psi _{1}^{\ast }\right) \psi _{2}\,.  \label{4}
\end{equation}

Therefore, $H_{0}$ is a symmetric operator \cite{4,5}. Besides, $H_{0}$ is a
positive operator.

In order to construct self-adjoint extensions of the operator $H_{0}$, one
should obtain the eigenfunctions $U^{\left( \pm \right) }$ of the adjoint
operator $H_{0}^{\ast }$ with the eigenvalues $\pm i$. In addition, $%
H_{0}^{\ast }$ is defined as the differential expression (\ref{3prime});
however, the domain is no longer restricted by the boundary values at $\rho
=0$ or at $\rho =\infty $. This implies that the function $U^{\left(
+\right) }$ (corresponding to the eigenvalue $+i$) must obey the equation%
\begin{equation}
\widetilde{H}U^{\left( +\right) }=iU^{\left( +\right) }  \label{5}
\end{equation}%
on the entire plane, except, perhaps, the point $\rho =0$. Let us search for
the eigenfunctions $U$ in the form%
\begin{equation}
U^{\left( +\right) }\left( \vec{\rho}\right) =e^{im\varphi }U_{m}^{\left(
+\right) }\left( \vec{\rho}\right) \,.  \label{6}
\end{equation}

Note that the function $e^{im\varphi }=\left( \frac{x+iy}{\rho }\right) ^{m}$
is infinitely differentiable everywhere except the origin. Therefore, the
action of the operator $\widetilde{H}$ on a function of the form (\ref{6})
(which ought to be known everywhere except the origin) does not require any
redefinition. The function $U_{m}^{\left( +\right) }$ satisfies the equation%
\begin{equation}
\left( -\partial _{\rho }^{2}-\frac{1}{\rho }\partial _{\rho }+\frac{\left(
m+\mu \right) ^{2}}{\rho ^{2}}\right) U_{m}^{\left( +\right) }\left( \vec{%
\rho}\right) =iU_{m}^{\left( +\right) }\left( \rho \right) \,.  \label{7}
\end{equation}

Eq. (\ref{7}) is the Bessel equation. Since (\ref{7}) must take place at $%
\rho \neq 0$, it does not require its redefinition neither in Cartesian nor
in cylindric coordinates. A solution of (\ref{7}) being square-integrable at
$\infty $ is given by%
\begin{equation}
U_{m}^{\left( +\right) }\left( \rho \right) =N_{m}^{-1}H_{m+\mu }^{\left(
1\right) }\left( e^{i\frac{\pi }{4}}\rho \right) \,,  \label{8}
\end{equation}%
where $N_{m}$ is a normalization factor (in case the function is
square-integrable). Function (\ref{8}) is square-integrable at zero (with
respect to the measure $dxdy\thicksim \rho d\rho $) only if $\left| m+\mu
\right| <1$. Therefore, the operator $H_{0}^{\ast }$ has two eigenfunctions
that belong to $L_{2}$ with the eigenvalue $i$ at a non-integer $\mu $ and
one eigenfunction $[...]$ an integer $\mu $. $H_{0}^{\ast }$ has the same
number of eigenfunctions with the eigenvalue $-i$. This follows from the
solution of an equation similar to (\ref{7}),%
\begin{equation}
U_{m}^{\left( -\right) }\left( \rho \right) =N_{m}^{-1}H_{m+\mu }^{\left(
2\right) }\left( e^{-i\frac{\pi }{4}}\rho \right) \,,  \label{8prime}
\end{equation}%
as well as from the fact that $H_{0}$ is a positive operator.

Therefore, $H_{0}$ is a symmetric operator with non-vanishing (and equal)
deficiency indices: $\left( 2,2\right) $ at $\Delta \mu \neq 0$ and $\left(
1,1\right) $ at $\Delta \mu =0$, where $\mu =-\left[ \mu \right] -\Delta \mu
$, $-\left[ \mu \right] $ being the integer part of $\mu $.

All essentially self-adjoint extensions $H_{\theta }$ of the operator $H_{0}$
are described as follows \cite{4,5}:

a) the domain is%
\begin{equation}
D_{\theta }=D_{0}+F+F^{\theta }\,,  \label{9}
\end{equation}%
where%
\begin{equation}
F=\left\{
\begin{array}{l}
A_{1}e^{i\left( \left[ \mu \right] +1\right) \varphi }N_{\left[ \mu \right]
+1}^{-1}H_{1-\Delta \mu }^{\left( 1\right) }\left( e^{i\frac{\pi }{4}}\rho
\right) +A_{2}e^{i\left[ \mu \right] \varphi }N_{\left[ \mu \right]
}^{-1}H_{\Delta \mu }^{\left( 1\right) }\left( e^{i\frac{\pi }{4}}\rho
\right) \,,\,\mathrm{at\,}\Delta \mu \neq 0 \\
Ae^{i\left[ \mu \right] \varphi }H_{0}^{\left( 1\right) }\left( e^{i\frac{%
\pi }{4}}\rho \right) \,,\,\mathrm{at\,}\Delta \mu =0\,,%
\end{array}%
\right.  \label{10}
\end{equation}

\begin{equation}
F^{\theta }=\left\{
\begin{array}{l}
B_{1}e^{i\left( \left[ \mu \right] +1\right) \varphi }N_{\left[ \mu \right]
+1}^{-1}H_{1-\Delta \mu }^{\left( 2\right) }\left( e^{-i\frac{\pi }{4}}\rho
\right) +B_{2}e^{i\left[ \mu \right] \varphi }N_{\left[ \mu \right]
}^{-1}H_{\Delta \mu }^{\left( 2\right) }\left( e^{-i\frac{\pi }{4}}\rho
\right) \,,\,\mathrm{at\,}\Delta \mu \neq 0 \\
Ae^{i\theta }e^{i\left[ \mu \right] \varphi }\rho H_{0}^{\left( 2\right)
}\left( e^{-i\frac{\pi }{4}}\rho \right) \,,\,\mathrm{at\,}\Delta \mu =0\,,%
\end{array}%
\right.  \label{10prime}
\end{equation}%
and $B_{i}=A_{j}\theta _{ji}$, $A_{i}$, $A$ are arbitrary complex numbers; $%
\theta _{ij}$ and $\theta $ are, respectively, an arbitrary (although fixed
for a given extension) unitary matrix and real number;

b)%
\begin{equation}
H_{\theta }D_{\theta }=H_{0}D_{0}+iF-iF^{\theta }\,.  \label{11}
\end{equation}

Self-adjoint extensions are given by the closure:\ $H_{\theta }:\overline{H}%
_{\theta }=H_{\theta }^{\ast }\,$. In addition, the domain of $\overline{H}%
_{\theta }$ is $\overline{D}_{\theta }=\overline{D}_{0}+F+F^{\theta }$,
where $\overline{D}_{\theta }$ is the domain of the operator $\overline{H}%
_{0}=H_{0}^{\ast \ast }$. The functions $\psi \subset \overline{D}_{0}$ are
non-singular.\footnote{%
The functions $\psi \subset \overline{D}_{0}$ become continuous after their
correction on the zero-measure set $\psi \left( 0\right) =0$; see \cite{5}.}

In order to select a self-adjoint extension, we require that the functions
from the domain of a ``physical'' operator should be nonsingular. This fixes
the extension in a unique way:

\begin{eqnarray}
\theta _{ij} &=&\left|
\begin{array}{cc}
\exp \left\{ -i\left( 1-\Delta \mu \right) \frac{\pi }{4}\right\} & 0 \\
0 & \exp \left\{ -i\Delta \mu \frac{\pi }{4}\right\}%
\end{array}%
\right| \,,\,\mathrm{for\;}\Delta \mu \neq 0\,,  \label{12} \\
\theta &=&0\,,\,\mathrm{for\,\;}\Delta \mu =0\,.  \label{13}
\end{eqnarray}

We must now obtain the complete system of generalized eigenfunctions for the
self-adjoint operator (which shall be denoted as $\mathcal{H}$),
corresponding to (\ref{12}) and (\ref{13}). Since the matrix $\theta _{ij}$
is diagonal, the subspaces of functions of the form $\left( \frac{x+iy}{\rho
}\right) ^{m}f\left( \rho \right) $ reduce $\mathcal{H}$, and we must find
generalized eigenfunctions of a ``nonsingular'' extension of Bessel's
differential expression on a semiaxis. This problem has been solved in \cite%
{6}, where all the inversion formulas related to Bessel's differential
expression of an arbitrary index $\nu \geq 0$ have been described. In the
case under consideration, the generalized eigenfunctions are given by%
\begin{equation}
e^{im\varphi }J_{\left| m+\mu \right| }\left( \sqrt{\lambda }\rho \right)
\,,\;m=0,\pm 1,\pm 2,\ldots ,  \label{14}
\end{equation}%
where $\lambda $ is an arbitrary real number, and the operator $\mathcal{H}$
proves to be positive. We note that in case $\Delta \mu =0$ the operator $%
\mathcal{H}$ also proves to be the only positive extension of the operator $%
H_{0}$.

Thus, it has been shown that under the physical condition of ``minimal
singularity'', i.e., the condition that the domain of the Hamiltonian should
be composed entirely of nonsingular functions, there exists a unique
self-adjoint operator $\mathcal{H}$ related to the differential expression (%
\ref{3}). It is natural to refer to the operator $\mathcal{H}$ as the
Hamiltonian of the problem in question. The complete system of generalized
eigenvectors of this operator is given by the set of functions (\ref{14}).

Let us now solve the scattering problem, i.e., construct a wave function $%
\psi _{k}\left( \vec{\rho}\right) $ subject to the Schr\"{o}dinger equation%
\begin{equation}
\mathcal{H}\psi _{k}\left( \vec{\rho}\right) =k^{2}\psi _{k}\left( \vec{\rho}%
\right) \,,\;k^{2}=E-k_{z}^{2}  \label{15}
\end{equation}%
(with $E$ being the total energy of the electron; $\vec{k}$ being the
projection of the total momentum of the electron on the $xy$ plane), as well
as to the following asymptotic condition:%
\begin{equation}
\psi _{k}\left( \vec{\rho}\right) \underset{\rho \rightarrow \infty }{%
\rightarrow }e^{i\vec{k}\vec{\rho}}+\frac{f\left( \varphi \right) }{\sqrt{%
2\pi k\rho }}e^{ik\rho -i\frac{\pi }{4}}\,.  \label{16}
\end{equation}

Condition (\ref{16}) implies that in case $\rho \rightarrow \infty $ the
wave function is a superposition of an incoming plain wave and a scattered
divergent cylindrical wave; $f\left( \varphi \right) $ is the scattering
amplitude.

Using the expansion%
\begin{equation}
e^{i\vec{k}\vec{\rho}}\rightarrow \frac{1}{\sqrt{2\pi k\rho }}\left[
e^{ik\rho -i\frac{\pi }{4}}\underset{m=-\infty }{\overset{\infty }{\sum }}%
e^{im\left( \varphi -\delta \right) }+e^{-ik\rho +i\frac{\pi }{4}}\underset{%
m=-\infty }{\overset{\infty }{\sum }}(-)^{m}e^{im\left( \varphi -\delta
\right) }\right]\,,  \label{17}
\end{equation}%
where $\delta $ is the angle between the vector $\vec{k}$ and the $x$ axis,
we obtain, due to the well-known asymptotic behavior of Bessel's functions,%
\footnote{%
Note that the wave function (\ref{18}) and the scattering amplitude (\ref{19}%
) are single-valued functions of the coordinates, whereas the wave function
of Aharonov and Bohm is not single-valued. This is related to the fact that,
instead of the asymptotic condition (\ref{16}), the authors of \cite{3} use
a condition that differs from (\ref{16}) by the multiplier $e^{i\mu \varphi }
$, which results in the non-single-valued character of the wave function in %
\cite{3}. The squared module of the scattering amplitude in \cite{3},
however, is identical with the squared module of function (\ref{19}).}%
\begin{equation}
\psi _{k}\left( \vec{\rho}\right) =e^{-i\mu \frac{\pi }{2}}\underset{m+\mu
\geq 0}{\sum }e^{im\left( \varphi -\delta \right) }i^{m}J_{m+\mu }\left(
k\rho \right) +e^{-i\mu \frac{\pi }{2}}\underset{m+\mu <0}{\sum }e^{im\left(
\varphi -\delta \right) }\left( -i\right) ^{m}J_{\left| m+\mu \right|
}\left( k\rho \right) \,.  \label{18}
\end{equation}

The scattering amplitude $\left( \varphi -\delta \neq 0\right) $ equals to%
\begin{equation}
f\left( \varphi \right) =e^{i\left( \left[ \mu \right] +\frac{1}{2}\right)
\left( \varphi -\delta \right) }\left( -\right) ^{\left[ \mu \right] }\frac{%
\sin ^{2}\Delta \mu \pi }{\sin ^{2}\left( \frac{\varphi -\delta }{2}\right) }%
\,.  \label{19}
\end{equation}

The cross-section of scattering equals to $\left( \varphi -\delta \neq
0\right) $%
\begin{equation}
\sigma \thicksim \left| f\left( \psi \right) \right| ^{2}=\frac{\sin
^{2}\Delta \mu \pi }{\sin ^{2}\left( \frac{\varphi -\delta }{2}\right) }\,,
\label{20}
\end{equation}%
which is identical with the result of \cite{3}. Therefore, it follows from (%
\ref{20}) that in the case of a quantized value of flow along an infinite,
infinitely thin solenoid ($\mu $ being integer) the scattering on the
potential is absent.

Let us make some more remarks concerning the case of an integer $\mu $.
Namely, we suppose $\mu =n$. The wave function then admits the representation%
\begin{equation}
\psi _{k}=\left( -\right) ^{n}e^{-in\left( \varphi -\delta \right) }e^{-i%
\vec{k}\vec{\rho}}\,.  \label{18prime}
\end{equation}%
(\ref{18prime}) has the following asymptotics:%
\begin{eqnarray}
&&\,\psi _{k}\left( \vec{\rho}\right) \rightarrow \frac{1}{\sqrt{2\pi \rho }}%
\left[ \left( -\right) ^{n}\delta \left( \psi -\delta \right) e^{ik\rho -i%
\frac{\pi }{4}}+\delta \left( \varphi -\delta -\pi \right) e^{-ik\rho +i%
\frac{\pi }{4}}\right]  \notag \\
&&\,=e^{i\vec{k}\vec{\rho}}+\left( \left( -\right) ^{n}-1\right) \delta
\left( \varphi -\delta \right) \frac{e^{ik\rho -i\frac{\pi }{4}}}{\sqrt{2\pi
k\rho }}\,.  \label{19prime}
\end{eqnarray}%
(\ref{19prime}) shows that scattering, in pure form, is absent only for even
$\mu $. For odd $\mu $, however, the electron ``senses'' the presence of the
string.

Another remark is as follows. The wave function (\ref{18prime}), as
mentioned in Introduction, contains the multiplier $e^{-in\varphi }$, being
a cause for a certain perplexity, which has been discussed in Introduction.
A solution of the mentioned paradox is given by the fact that for integer $%
\mu $ the potential $A_{k}^{\left( c\right) }$ can actually be transformed
out of the Schr\"{o}dinger equation with the help of a \emph{gauge
transformation}, which, however, is not a \emph{gradient transformation}.
Namely, the potential $A_{k}^{\left( c\right) }$ can be represented in the
form (for integer $\mu =n$)%
\begin{equation}
A_{k}^{\left( c\right) }=-ie^{in\varphi }\partial _{k}e^{-in\varphi }\equiv
-i\left( \frac{x+iy}{\rho }\right) ^{n}\partial _{k}\left( \frac{x-iy}{\rho }%
\right) ^{n}\,.  \label{21}
\end{equation}

For non-integer $\mu $, such a representation \emph{does not take place}!

It can be readily verified that (\ref{21}) holds true in the entire space,
except, perhaps, the origin. Representation (\ref{21}) turns out to be
sufficient to eliminate the potential from the Schr\"{o}dinger equation with
the help of the gauge transformation%
\begin{equation}
A_{k}\rightarrow A_{k}-iU^{-1}\partial _{k}U\,,\;\psi \rightarrow U\psi \,.
\label{22}
\end{equation}

In order that the transformation of $A_{k}$ be a gradient one, it is
necessary to impose the subsidiary condition of a continuous
differentiability of $\ln U$, which is not necessarily fulfilled in the
general case. This is precisely what happens in the problem under
consideration. Representation (\ref{21}) is not reduced to a pure gradient,
since in this case $\ln U=-i\varphi $ is a discontinuous function (for
instance, along the line $y=0$, $x>0$).

For the sake of completeness, one should investigate the point $\rho =0$
more accurately. We will, however, restrict ourselves to the above
qualitative remark.

\section{A thin solenoid as a limit of a thick solenoid}

Let us replace the potential $A_{k}^{\left( c\right) }$ in expression (\ref%
{3}) by a regularized potential $A_{k}^{\left( R\right) }$:%
\begin{eqnarray}
&&A_{k}^{\left( R\right) }\left( \vec{r}\right) =A_{k}^{\left( c\right)
}\left( \vec{r}\right) f_{a}\left( \rho \right) \,,  \label{23} \\
&&f_{a}\left( \rho \right) =1\,,\;\rho \geq a\,;\;f_{a}\left( 0\right) =0\,;
\label{24}
\end{eqnarray}%
besides, $f_{a}$ is continuous at the point $\rho =a$. Note that the flow of
magnetic field corresponding to the potential (\ref{23}) is equal to $\mu $.

Let us select the eigenfunctions of $H$ as solutions being regular at zero,
which is usually postulated on the basis of physical reasons.\footnote{%
From the mathematical viewpoint, one can say that the Hamiltonian, defined
as the differential expression (\ref{3}) with potential (\ref{23}) and a
domain consisting of the set of all the twice-differentiable functions with
a compact support, is an essentially self-adjoint operator. This operator is
also a unique self-adjoint operator with a ``nonsingular'' domain
corresponding to the differential expression (\ref{3}) with potential (\ref%
{23}).} These functions are%
\begin{equation}
U_{k}^{\left( m\right) }\left( \vec{\rho}\right) =\left\{
\begin{array}{l}
c_{m}F_{m,k}\left( \rho \right) e^{im\varphi }\,,\;\rho \leq a\,, \\
c_{m}\left( A_{m}J_{\left| m+\mu \right| }\left( k\rho \right)
+B_{m}N_{\left| m+\mu \right| }\left( k\rho \right) \right) e^{im\varphi
}\,,\;\rho \geq a\,,%
\end{array}%
\right.  \label{25}
\end{equation}

The function $F_{m,k}\left( \rho \right) $ is a solution, regular at zero
(normalized as $\rho ^{|m|}$ when $\rho \rightarrow 0$), of the following
equation:%
\begin{equation}
\left[ -\partial _{\rho }^{2}-\frac{1}{\rho }\partial _{\rho }+\frac{1}{\rho
^{2}}\left( f_{a}\left( \rho \right) \mu ^{2}+2m\mu f_{a}\left( \rho \right)
+m^{2}\right) \right] F_{m,k}\left( \rho \right) =k^{2}F_{m,k}\left( \rho
\right) \,.  \label{26}
\end{equation}%
The coefficients $c_{m}$ are arbitrary, whereas the coefficients $A_{m}$ and
$B_{m}$ are chosen from the condition that $U_{k}^{\left( m\right) }\,$and
their first derivatives be continuous at the point $\rho =a$:%
\begin{eqnarray}
A_{m} &=&\frac{\pi a}{2}\left[ F_{m,k}\left( a\right) N_{\left| m+\mu
\right| }^{\prime }\left( ka\right) -F_{m,k}^{\prime }\left( a\right)
N_{\left| m+\mu \right| }\left( ka\right) \right] \,,  \label{27} \\
B_{m} &=&\frac{\pi a}{2}\left[ F_{m,k}^{\prime }\left( a\right) J_{\left|
m+\mu \right| }\left( ka\right) -F_{m,k}\left( a\right) J_{\left| m+\mu
\right| }^{\prime }\left( ka\right) \right] \,.  \label{27prime}
\end{eqnarray}

The solution $\Psi _{k}\left( \vec{\rho}\right) $ of the Schr\"{o}dinger
equation with asymptotics (\ref{16}) is given by%
\begin{eqnarray}
\Psi _{k}\left( \vec{\rho}\right) &=&\psi _{k}\left( \vec{\rho}\right) -i%
\underset{m}{\sum }e^{im\left( \varphi -\delta \right) +im\pi -i\left| m+\mu
\right| \frac{\pi }{2}}\frac{b_{m}}{1+ib_{m}}H_{\left| m+\mu \right|
}^{\left( 1\right) }\left( k\rho \right) \,,\;\rho \geq a\,,  \label{28} \\
\Psi _{k}\left( \vec{\rho}\right) &=&\underset{m}{\sum }\frac{\left(
-\right) ^{m}e^{im\left( \varphi -\delta \right) -i\left| m+\mu \right|
\frac{\pi }{2}}}{1+ib_{m}}\frac{F_{m,k}\left( \rho \right) }{A_{m}}\,,\;\rho
\leq a\,.  \label{28prime}
\end{eqnarray}

The corresponding scattering amplitude $\phi _{k}$ equals to ($\varphi
-\delta \neq 0$)%
\begin{equation}
\phi _{k}=f\left( \varphi \right) -2i\underset{m}{\sum }\frac{b_{m}}{1+ib_{m}%
}\left( -\right) ^{m}e^{im\left( \varphi -\delta \right) -i\left| m+\mu
\right| \pi }\,.  \label{29}
\end{equation}

In (\ref{28}) and (\ref{29}), the functions $\psi _{k}\left( \vec{\rho}%
\right) $ and $f\left( \varphi \right) $ are given by formulas (\ref{18}), (%
\ref{19}) and present solutions of the problem for an infinitely thin
solenoid; the coefficients $b_{m}$ are given by%
\begin{equation}
b_{m}=\frac{B_{m}}{A_{m}}\,.  \label{30}
\end{equation}

Using the notation%
\begin{equation}
F_{m,k}\left( \rho \right) =\rho ^{\left| m\right| }\widetilde{F}%
_{m,k}\left( \rho \right) \,,\;\widetilde{F}_{m,k}\left( 0\right) =1\,,
\label{31}
\end{equation}%
one can present $b_{m}$ as follows:%
\begin{equation}
b_{m}=\frac{J_{\left| m+\mu \right| }\left( ka\right) }{\left( \left|
m\right| +\left| m+\mu \right| \right) N_{\left| m+\mu \right| }\left(
ka\right) }\cdot \frac{\left| m+\mu \right| -\left| m\right| -\frac{a%
\widetilde{F}_{m,k}^{\prime }\left( a\right) }{\widetilde{F}_{m,k}\left(
a\right) }-\frac{akJ_{\left| m+\mu \right| +1}\left( ka\right) }{J_{\left|
m+\mu \right| }\left( ka\right) }}{1+\frac{1}{\left| m\right| +\left| m+\mu
\right| }\left[ \frac{a\widetilde{F}_{m,k}^{\prime }\left( a\right) }{%
\widetilde{F}_{m,k}\left( a\right) }-\frac{akN_{\left| m+\mu \right|
-1}\left( ka\right) }{N_{\left| m+\mu \right| }\left( ka\right) }\right] }\,.
\label{30prime}
\end{equation}

Let us take into account an estimation for the Bessel function \cite{8},
which is implied by its representation as a series:%
\begin{eqnarray}
&&J_{\nu }\left( z\right) =\frac{z^{\nu }}{2^{\nu }\Gamma \left( \nu
+1\right) }\left( 1+\theta \right) \,,  \label{32} \\
&&\left| \theta \right| <\exp \left\{ \frac{\left| z\right| ^{2}}{4\left(
1+\nu _{0}\right) }\right\} -1\,,\;\nu _{0}=\min \left( \left| \nu +1\right|
,\left| \nu +2\right| ,\ldots \right) \,.  \label{32prime}
\end{eqnarray}

In a similar way, one can obtain an estimation for $N_{\nu }$, $\nu \geq 0$:%
\begin{eqnarray}
N_{\nu }\left( z\right) &=&-\frac{2^{\nu }\Gamma \left( \nu \right) }{\pi
z^{\nu }}\left( 1+\theta _{1}\right) \,,\;\nu \neq 0\,,  \label{33} \\
N_{0}\left( z\right) &=&\frac{2}{\pi }\ln \frac{z}{2}\left( 1+\theta
_{2}\right) \,,  \label{34}
\end{eqnarray}%
where $\theta _{1}$ and $\theta _{2}$ are bounded (uniformly in $m$, $\nu
=\left| m+\mu \right| $, with a fixed $\mu $) functions in any bounded
region $z$; $\theta _{i}\rightarrow 0$ as $\left| z\right| \rightarrow 0$,
while $\left| \theta _{1}\right| <C\left| z\right| ^{2}$ at $\nu >1$, the
constant $C$ being independent of $m$, $\left| \theta _{1}\right| \thicksim
\left| z\right| ^{2}\ln \left| z\right| $ at $\nu =1$, $\left| \theta
_{1}\right| \thicksim \left| z\right| ^{2\nu }$, at $0<\nu <1$, $\left|
\theta _{2}\right| \thicksim 1/\ln \left| z\right| $.

From (\ref{30}), (\ref{30prime}), (\ref{27}), (\ref{27prime}), it is easy to
see that all $b_{m}\rightarrow 0$ as $a\rightarrow 0$. Suppose, furthermore,
that the quantities%
\begin{equation}
\frac{a\widetilde{F}_{m,k}^{\prime }\left( a\right) }{\widetilde{F}%
_{m,k}^{\prime }\left( a\right) }  \label{35}
\end{equation}%
are uniformly bounded in $m$ and $a$ at $ka<\varepsilon $ for a sufficiently
small $\varepsilon $. Then (\ref{30prime}) and (\ref{32})--(\ref{34}) imply
an estimation\footnote{%
(\ref{36}) gives a correct estimation for $m+\mu \neq 0$. For $m+\mu =0$, we
have $b_{\left[ \mu \right] }\thicksim 1/\ln a$. It is important, however,
that $b_{\left[ \mu \right] }\rightarrow 0$ as $a\rightarrow 0$.} for $b_{m}$%
:%
\begin{equation}
b_{m}=\frac{\pi \left( ka\right) ^{2\left| m+\mu \right| }}{2^{2\left| m+\mu
\right| }\left( \left| m\right| +\left| m+\mu \right| \right) \Gamma \left(
\left| m+\mu \right| \right) \Gamma \left( \left| m+\mu \right| +1\right) }%
Q_{m,k}\left( a\right) \,,  \label{36}
\end{equation}%
where the quantities $Q_{m,k}\left( a\right) $ are uniformly bounded in $m$
and $a$ for $ka<\varepsilon $.

With the help of (\ref{36}), we easily find%
\begin{equation}
\left| \phi _{k}\left( \varphi \right) -f\left( \varphi \right) \right| \leq
2\underset{\left| m+\mu \right| \leq 1}{\sum }\left| b_{m}\right|
+a^{2}\theta \left( a\right) \,,\;\varphi -\delta \neq 0\,,  \label{37}
\end{equation}%
$\theta \left( a\right) $ being a bounded function as $a\rightarrow 0$.

Therefore, $\phi _{k}\left( \bar{\varphi}\right) \rightarrow \bar{f}\left(
\varphi \right) $ as $a\rightarrow 0$ for all $\varphi $, $\varphi -\delta
\neq 0$, at any value of $\mu $.

It remains to prove that the values $a\widetilde{F}_{m,k}^{\prime }\left(
a\right) /\widetilde{F}_{m,k}\left( a\right) $ are uniformly bounded in $m$
and $a$ at $ka<\varepsilon $. This is done in Appendix B.

Let us examine the quantity $\left| \Psi _{k}-\psi _{k}\right| $. Using
estimations (\ref{33}), (\ref{34}) and (\ref{36}), we obtain, for any
bounded region on the plane $xy$ at $\rho \geq a$:%
\begin{eqnarray}
\left| \Psi _{k}-\psi _{k}\right| &\thicksim &\left| \underset{\left| m+\mu
\right| <1}{\sum }b_{m}N_{\left| m+\mu \right| }\left( k\rho \right) \right|
+\underset{\left| m+\mu \right| \geq 1}{\sum }\left( \frac{ka}{2}\right)
^{\left| m+\mu \right| }\frac{1}{\left( \left| m\right| +\left| m+\mu
\right| \right) \Gamma \left( \left| m+\mu \right| +1\right) }\left( \frac{a%
}{\rho }\right) ^{\left| m+\mu \right| }  \notag \\
&\thicksim &\left| \underset{\left| m+\mu \right| <1}{\sum }b_{m}N_{\left|
m+\mu \right| }\left( k\rho \right) \right| +ka\phi \left( \frac{a}{\rho }%
\right) \thicksim \mathrm{const\,}\cdot a^{\Delta \mu }+\mathrm{const\,}%
\cdot a^{1-\Delta \mu }+\mathrm{const\,}\cdot a\,,  \label{38}
\end{eqnarray}%
where $\phi \left( \frac{a}{\rho }\right) $ is a bounded (in a bounded part
of the plane, at $\rho \geq a$) function is uniform in $a$ as $a\rightarrow
0 $.

Let us further use estimations (\ref{33}), (\ref{34}), (\ref{36}), (\ref{41}%
) and obtain for $\rho \leq a$:%
\begin{eqnarray}
&&\left| \Psi _{k}-\psi _{k}\right| \thicksim \left| \underset{\left| m+\mu
\right| }{\sum }J_{\left| m+\mu \right| }\left( k\rho \right) \right| +%
\mathrm{const}\cdot a\thicksim \mathrm{const}\cdot a^{\Delta \mu }+\mathrm{%
const}\cdot a^{1-\Delta \mu }+\mathrm{const}\cdot a\,,  \label{39} \\
&&\left| \Psi _{k}\left( \vec{\rho}\right) \right| \thicksim \underset{m}{%
\sum }\frac{1}{\left| m\right| +\left| m+\mu \right| }\cdot \frac{1}{%
N_{\left| m+\mu \right| }\left( ka\right) }\left( \frac{\rho }{a}\right)
^{_{\left| m\right| }}\rightarrow 0\,,\;\mathrm{as\,}a\rightarrow 0\,.
\label{39prime}
\end{eqnarray}

Estimations (\ref{38}), (\ref{39}) and (\ref{39prime}) lead to the following
result:\footnote{%
The first term in (\ref{38}) is of order $a^{\Delta \mu }\left( \frac{a}{%
\rho }\right) ^{\Delta \mu }+a^{1-\Delta \mu }\left( \frac{a}{\rho }\right)
^{1-\Delta \mu }\thicksim a^{\Delta \mu }+a^{1-\Delta \mu }$ at $\Delta \mu
\neq 0$. However, at $\Delta \mu =0$ ($\mu =-n$) the first term in (\ref{38}%
) is of order $\ln \rho /\ln a$. At a fixed $\rho $, it tends to zero as $%
a\rightarrow 0$; however, the same term is $\thicksim 1$ at $\rho =a$.}

a)\ for $\Delta \mu \neq 0$, $\Psi _{k}\left( \vec{\rho}\right) $ converges
as $a\rightarrow 0$ to the function $\psi _{k}\left( \vec{\rho}\right) $
uniformly in any bounded part of the plane $xy$;

b)\ at $\Delta \mu =0$ this convergence is uniform only if $\rho >\delta $
for an arbitrarily small but fixed $\delta $. This can be observed, in
particular, from the fact that $\Psi _{k}\left( 0\right) \thicksim a^{n}$
and $\psi _{k}\left( 0\right) \thicksim 1$. However, the quantity $\left|
\Psi _{k}\left( \vec{\rho}\right) \right| ^{2}dxdy\thicksim \left| \Psi
_{k}\left( \vec{\rho}\right) \right| ^{2}\rho d\rho $, being the
probability, ``uniformly'' converges to the probability $\left| \psi
_{k}\left( \vec{\rho}\right) \right| ^{2}dxdy$ in the sense that $%
\int_{\Delta }\left( \left| \Psi _{k}\right| ^{2}-\left| \psi _{k}\right|
^{2}\right) dxdy$ tends to zero as $a\rightarrow 0$ uniformly with respect
to an arbitrary choice of the integration region $\Delta \subset D$, $D$
being an arbitrary bounded region of the plane $xy$.

Thus, it has been proved that the probability distribution in the problem of
electron scattering by a solenoid of radius $a$, with any distribution of
magnetic field inside of it, converges uniformly as $a\rightarrow 0$ and
fixed flow of magnetic field $\mu $, in any limited region of the plane (in
the integral sense that has been explained above) to the probability
distribution in the problem of electron scattering by an infinitely thin
solenoid. The scattering amplitude corresponding to a thick solenoid
converges as $a\rightarrow 0$ to the scattering amplitude corresponding to a
thin solenoid uniformly in the region $\varepsilon <\varphi -\delta <2\pi
-\varepsilon $ for any fixed $\varepsilon >0$.

\section{Electron scattering by the field of a semi-infinite thin solenoid}

As a matter of fact, we shall examine, in the first place, the case of two
semi-infinite infinitely thin solenoids, with the corresponding potential%
\begin{equation}
A_{k}^{(S)}\left( \vec{r}\right) =-\mu \frac{\varepsilon _{3ki}r_{i}}{r}%
\left( \frac{1}{r-z}-\frac{1}{r+z}\right) =-\mu _{1}\frac{\varepsilon
_{3ki}r_{i}}{r}\frac{z}{\rho ^{2}}\,,\mu _{1}=2\mu \,.  \label{40}
\end{equation}

Potential (\ref{40}) corresponds to the potential of a monopole in
Schwinger's formulation \cite{1}.

The Hamiltonian of the problem is given by%
\begin{eqnarray}
H &=&\left( -i\partial _{k}+A_{k}^{(S)}\right) +V\left( r\right) =-\frac{1}{%
r^{2}}\partial _{r}\left( r^{2}\partial _{r}\right) +\frac{1}{r^{2}}J^{2}-%
\frac{\mu _{1}^{2}}{r^{2}}+V\left( r\right) \,,  \label{41} \\
J^{2} &=&\frac{1}{\sin \theta }\partial _{\theta }\left( \sin \theta
\partial _{\theta }\right) +\frac{1}{\sin ^{2}\theta }\left[ -\partial
_{\theta }^{2}-2i\mu _{1}\cos \theta \partial _{\theta }+\mu _{1}^{2}\right]
\,.  \label{42}
\end{eqnarray}

We have added to the Hamiltonian the potential $V$ (which in the case of
purely electromagnetic interaction between the electron and monopole is to
be set equal to zero).

We shall now construct a self-adjoint operator corresponding to the
differential expression (\ref{42}). Let us define the operator $J_{0}^{2}$
as the differential operator (\ref{42}) with the following domain $D_{0}$:
1)\ the functions $\psi \left( \theta ,\varphi \right) $ from $D_{0}$ must
be twice-differentiable; 2)\ $\psi $ and $\partial _{\varphi }\psi $ must be
periodic in $\varphi $ with the period $2\pi $; 3)\ $\psi =0$ at $\theta
<\varepsilon $ and $\theta >\pi -\varepsilon $ for a certain $\varepsilon $,
where $\varepsilon $ may be different for different functions. Property (2)
is necessary due to physical reasons, as well as due to the fact that the
Hamiltonian be defined on these functions in Cartesian coordinates in a
natural way, since the singularities of the Hamiltonian are located only on
the $z$ axis. Note also that the functions with properties (1) and (2) form
the domain of a self-adjoint operator corresponding to the differential (in $%
\varphi $) expression of second order with constant coefficients, on the
segment $\left[ 0,2\pi \right] $. The eigenfunctions of this operator are $%
e^{-im\varphi }$, $m=0,\pm 1,\pm 2\ldots $, ..., which are usually selected
due to physical reasons.

$D_{0}$ is dense in the space $L_{2}$ of functions on a unit sphere, so that
$J_{0}^{2}$ is a symmetric operator. Besides, $J_{0}^{2}$ is a positive
operator. Let us find the eigenfunctions $U$ of the operator $\left(
J_{0}^{2}\right) ^{\ast }$. Representing the function $U$ in the form $%
U\left( \theta ,\varphi \right) =e^{-im\varphi }U_{m}\left( \theta \right) $%
, we find that $U_{m}\left( \theta \right) $ satisfies the equation%
\begin{equation}
\frac{1}{\sin ^{2}\theta }\left[ -\sin \theta \partial _{\theta }\left( \sin
\theta \partial _{\theta }\right) +m^{2}+\mu _{1}^{2}-2m\mu _{1}\cos \theta %
\right] U_{m}\left( \theta \right) =\lambda U_{m}\left( \theta \right) \,.
\label{43}
\end{equation}

Let us recall that the eigenfunctions of the operator $\left(
J_{0}^{2}\right) ^{\ast }$ are not subject to any boundary conditions \cite%
{4,5}. Eq. (\ref{43}) has an exact solution, and the eigenfunctions $%
U_{m,l}\left( \theta ,\varphi \right) $ are given by%
\begin{eqnarray}
U_{m,l}^{\left( 1\right) } &=&N_{m,l}^{-1}e^{-im\varphi }\left( \frac{1-t}{2}%
\right) ^{\frac{\alpha }{2}}\left( \frac{1+t}{2}\right) ^{\frac{\beta }{2}%
}F\left( -l,l+\alpha +\beta +1;1+\alpha ;\frac{1-t}{2}\right) \,,  \notag \\
U_{m,l}^{\left( 2\right) } &=&M_{m,l}^{-1}e^{-im\varphi }\left( \frac{1-t}{2}%
\right) ^{\frac{\alpha }{2}}\left( \frac{1+t}{2}\right) ^{\frac{\beta }{2}%
}F\left( -l,l+\alpha +\beta +1;1+\beta ;\frac{1+t}{2}\right) \,,  \label{44}
\end{eqnarray}%
where%
\begin{eqnarray}
\lambda &=&\left( l+\frac{\alpha +\beta }{2}\right) \left( l+\frac{\alpha
+\beta }{2}+1\right) \,,  \notag \\
\alpha &=&\left| -m+\mu _{1}\right| \,,\;\beta =\left| m+\mu _{1}\right|
\,,\;t=\cos \theta \,.  \label{45}
\end{eqnarray}%
$F\left( a,b;c;x\right) $ is a hypergeometric function \cite{9}; $N$ and $M$
are normalization coefficients (provided that the corresponding functions
are normalizable). An analysis of the hypergeometric function $F\left(
a,b;c;x\right) $ as $x\rightarrow 1$ shows that for a complex $\lambda $ the
functions $U_{m,l}^{\left( 1\right) }\subset L_{2}$ only on condition that $%
\left| m+\mu _{1}\right| <1$, whereas $U_{m,l}^{\left( 2\right) }\subset
L_{2}$ only on condition that $\left| -m+\mu _{1}\right| <1$. Thus, for
every complex-valued $\lambda $ the operator $\left( J_{0}^{2}\right) ^{\ast
}$ has $4$ eigenfunctions from $L_{2}$ for $\Delta \mu _{1}\neq 0$ ($\mu
_{1}=-\left[ \mu _{1}\right] -\Delta \mu _{1}$) and $2$ such functions for $%
\Delta \mu _{1}=0$.

All essentially self-adjoint extensions $J_{T}^{2}$ of the operator $%
J_{0}^{2}$ are described \cite{4,5} by the following domains $D_{T}$:%
\begin{equation}
D_{T}=D_{0}+A_{i}U_{i}+A_{i}T_{ij}\overline{U}_{j}\,,  \label{46}
\end{equation}%
where $U_{i}$ and $\overline{U}_{i}$ are functions of the form (\ref{44})
for some fixed complex $l$:%
\begin{eqnarray}
&&U_{1,2}=U_{mi,l}^{\left( 2\right) }\,,\;m_{1}=-\left[ \mu _{1}\right]
\,,\;m_{2}=-\left[ \mu _{1}\right] -1\,,  \notag \\
&&U_{3,4}=U_{mi,l}^{\left( 1\right) }\,,\;m_{3}=\left[ \mu _{1}\right]
\,,\;m_{4}=\left[ \mu _{1}\right] +1\,,\;\Delta \mu _{1}>0\,,  \label{47} \\
&&  \notag \\
&&U_{1}=U_{\left[ \mu _{1}\right] ,l}^{\left( 1\right) }\,,\;U_{2}=U_{-\left[
\mu _{1}\right] ,l}^{\left( 2\right) }\,,\;\Delta \mu _{1}=0\,.
\label{47prime}
\end{eqnarray}%
The functions $\overline{U}_{i}$ are obtained form the functions $U_{i}$ by
the change $l\rightarrow l^{\ast }$, $T_{ij}$ being an arbitrary (however
fixed for a given extension) unitary matrix of dimension $4\times 4$ for $%
\Delta \mu _{1}\neq 0$ and $2\times 2$ for $\Delta \mu _{1}=0$.

The requirement of ``minimal singularity'' for the self-adjoint extension
(which we further denote as $J^{2}$) of the operator $J_{0}^{2}$, that is,
the condition\ that the domain of $J^{2}$ should contain only non-singular
functions uniquely determines the extension (the matrix $T$).

We do not consider the case $\Delta \mu _{1}\neq 0$. Let us only observe
that all the functions from the domain of $J^{2}$ turn to zero along the $z$
axis (at least, as $\theta ^{\Delta \mu _{1}}$ or $\theta ^{1-\Delta \mu
_{1}} $, when $\theta \rightarrow 0$, and $\left( \pi -\theta \right)
^{^{\Delta \mu _{1}}}$ or $\left( \pi -\theta \right) ^{^{1-\Delta \mu
_{1}}} $, when $\theta \rightarrow \pi $). For any other extension of the
operator $J_{0}^{2}$ they either turn to zero or are singular on the $z$
axis. In any case, this means that for $\Delta \mu _{1}\neq 0$ the electron
experiences scattering by the string.

We shall now examine the case $\Delta \mu _{1}$ ($\mu _{1}=-n$). The matrix $%
T$ for a ``non-singular'' extension is determined, as has been observed, in
a unique way:%
\begin{equation}
T_{kj}=\frac{n+i}{n-i}\delta _{kj}\,.  \label{48}
\end{equation}

It is easy to see that the eigenfunctions of the operator $J^{2}$ from $%
L_{2} $ for this extension are given by (\ref{44}), where $l$ is allowed to
take only positive integer values (for any other real $l$ functions (\ref{48}%
) are singular at $\theta =0$ or at $\theta =\pi $):%
\begin{equation}
l=0,1,2,\ldots  \label{49}
\end{equation}%
(another possible set $l+\alpha +\beta +1=0,-1,-2,\ldots $ gives the same
set of eigenfunctions and eigenvalues). Therefore, $J^{2}$ is a positive
operator.

For integer $l$, the hypergeometric functions%
\begin{equation*}
F\left( -l,l+\alpha +\beta +1;1+\alpha ;\frac{1-t}{2}\right) \,,\;F\left(
-l,l+\alpha +\beta +1;1+\beta ;\frac{1+t}{2}\right)
\end{equation*}
are proportional to each other, as well as to the Jacobi polynomials \cite{9}
$P_{n}^{\alpha ,\beta }\left( t\right) $ (thus, for integer $l$ the
solutions $U_{m,l}^{\left( 1\right) }$ and $U_{m,l}^{\left( 2\right) }$
coincide; the second independent solution in this case is singular at one of
the ends $\theta =0$ or $\theta =2\pi $). Since in the case of fixed $\alpha
$ and $\beta $ the Jacobi polynomials form a complete orthonormal system in
the space $L_{2}$ of functions on the segment $\left[ -1,1\right] $, with a
scalar product determined by the weight $\left( 1-t\right) ^{-\alpha }\left(
1+t\right) ^{\beta }$, we find as a result that the eigenfunctions $U_{m,l}$
of the operator $J^{2}$,%
\begin{equation}
U_{m,l}=N_{m.l}^{-1}e^{-im\varphi }\left( \frac{1-t}{2}\right) ^{\frac{%
\alpha }{2}}\left( \frac{1+x}{2}\right) ^{\frac{\beta }{2}}F\left(
-l,l+\alpha +\beta +1;1+\alpha ;\frac{1-t}{2}\right) \,,  \label{44prime}
\end{equation}%
form a complete system of vectors in the space of functions on a unit sphere.

This means that functions (\ref{44prime}) form a complete system of
(generalized)\ eigenvectors of the operator $J^{2}$, whose spectrum is
discrete and consists of the points%
\begin{equation}
\lambda _{L}=L\left( L+1\right) \,,\;L\geq \left| \mu _{1}\right| \,,
\label{50}
\end{equation}%
with each value $\lambda _{L}$ being $\left( 2L+1\right) $-times degenerate
in $m\,\;\left( -L\leq m\leq L\right) $.

Functions (\ref{44prime}) coincide (up to a phase multiplier) with the
generalized spherical functions $T_{L}^{m,\mu _{1}}\left( \varphi ,\theta
,0\right) $ (see \cite{10}). The angular form of solution (\ref{44prime})
and its relation with the group of rotations has been known for a long time
(see \cite{11,12} as well as the discussion in \cite{1,13}), and we will not
dwell on this subject. Let us only note, once again, that, as has been
proved, the operator $J^{2}$, determined by its eigenvectors and spectrum (%
\ref{44prime}) (\ref{50}), is the only self-adjoint extension (of the
differential expression (\ref{42})) with a ``non-singular'' domain.

Let us now construct solutions of the Schr\"{o}dinger equation with
Hamiltonian (\ref{41}) (where $J^{2}$ is to be understood as the chosen
self-adjoint extension) and with the asymptotic condition%
\begin{equation}
\psi _{k}\left( \vec{r}\right) \rightarrow e^{i\vec{k}\vec{r}}+\frac{f\left(
\theta ,\varphi \right) }{r}e^{ikr}\,,  \label{51}
\end{equation}%
where $E=k^{2}$ is the energy of the electron.

The general solution of the Schr\"{o}dinger equation has the form%
\begin{equation}
\psi _{k}\left( \vec{r}\right) =\underset{m,L}{\sum }C_{m,L}T_{L}^{m,\mu
_{1}}\left( \varphi ,\theta ,0\right) R_{L,k}\left( r\right) i^{L}\,,
\label{52}
\end{equation}%
where $C_{m,L}$ are arbitrary coefficients, while the functions $R_{L,k}$
obey the equation%
\begin{eqnarray}
&&\left[ \frac{1}{r^{2}}\partial _{r}\left( r^{2}\partial _{r}\right) -\frac{%
L\left( L+1\right) }{r^{2}}-V_{1}\left( r\right) +k^{2}\right] R_{L,k}\left(
r\right) =0\,,  \label{53} \\
&&V_{1}\left( r\right) =V\left( r\right) -\frac{\mu _{1}^{2}}{r^{2}}
\label{54}
\end{eqnarray}%
and have the following asymptotics as $r\rightarrow \infty $:%
\begin{equation}
R_{L,k}\left( r\right) \rightarrow \frac{1}{kr}\sin \left( kr-\frac{\pi }{2}%
L+\delta _{L}\right) \,.  \label{55}
\end{equation}

Since the functions $T_{L}^{m,\mu _{1}}$ form a complete system on a sphere,
the asymptotics of the function $e^{i\vec{k}\vec{r}}$ can be presented in
the form\footnote{%
The functions $T_{L}^{m,\mu _{1}}$ are normalized on a sphere by $\sqrt{%
\frac{4\pi }{2L+1}}$ exactly as the usual Legendre polynomials.} (see
Appendix C):%
\begin{eqnarray}
&&e^{i\vec{k}\vec{r}}\underset{r\rightarrow \infty }{\rightarrow }\frac{1}{%
2ikr}\left[ e^{ikr}\underset{m,L}{\sum }T_{L}^{\ast m,\mu _{1}}\left(
\varphi _{k},\theta _{k},0\right) T_{L}^{m,\mu _{1}}\left( \varphi ,\theta
,0\right) \left( 2L+1\right) \right.  \notag \\
&&-\left. e^{-ikr}\underset{m,L}{\sum }T_{L}^{\ast m,\mu _{1}}\left( \pi
+\varphi _{k},\pi -\theta _{k},0\right) T_{L}^{m,\mu _{1}}\left( \varphi
,\theta ,0\right) \left( 2L+1\right) \right] \,,  \label{56}
\end{eqnarray}%
where $\varphi _{k}$, $\theta _{k}$ are the angular coordinates of the
vector $\vec{k}$.

As a result, we find that the coefficients $C_{m,L}$ have the form%
\begin{equation}
C_{m,L}=\left( -\right) ^{L}\left( 2L+1\right) e^{i\delta _{L}}T_{L}^{\ast
m\mu _{1}}\left( \pi +\varphi _{k},\pi -\theta _{k},0\right) \,.  \label{57}
\end{equation}

Let us take into account the following property of the functions $%
T_{L}^{m,\mu _{1}}$:%
\begin{equation}
T_{L}^{\ast m,\mu _{1}}\left( \varphi ,\theta ,\psi \right) =T_{L}^{m,\mu
_{1}}\left( \pi -\psi ,\theta ,\pi -\varphi \right) \,,  \label{58}
\end{equation}%
and the addition formula for spherical functions \cite{10}. Then, the wave
function can be presented in the form\footnote{%
The functions $T_{L}^{m,\mu }$ have the form \cite{10} $T_{L}^{m,\mu }\left(
\varphi ,\theta ,\psi \right) =e^{-im\varphi -i\mu \psi }P_{L}^{m,\mu
}\left( \theta \right) $.}

\begin{eqnarray}
&&\psi _{k}\left( \vec{r}\right) =e^{i\mu _{1}\Omega _{1}-i\mu _{1}\pi }%
\underset{L}{\sum }\left( 2L+1\right) i^{L}P_{L}^{-\mu _{1},\mu _{1}}\left(
\theta _{kr}\right) e^{i\delta _{L}}R_{L,k}\left( r\right) \,,  \label{59} \\
&&\tan \frac{\Omega _{1}}{2}=\frac{\sin \frac{\theta +\theta _{k}}{2}}{\sin
\frac{\theta -\theta _{k}}{2}}\tan \frac{\varphi -\varphi _{k}}{2}\,,\;\cos
\theta _{kr}=\frac{\vec{k}\vec{r}}{kr}\,.  \label{60}
\end{eqnarray}

For convenience, let us present the expression (\ref{59}) for the wave
function in the coordinate system related to the old coordinate system by
the Euler angles $\left( \frac{\pi }{2}-\varphi _{k}\,,\theta _{k}\,,\alpha +%
\frac{\pi }{2}\right) $. In the new coordinate system, the $z$ axis is
directed along the vector $\vec{k}$, whereas the potential has the form%
\begin{eqnarray}
&&\vec{A}^{\left( S,n\right) }\left( \vec{r}\right) =\mu _{1}\frac{\vec{n}%
\times \vec{r}}{r}\cdot \frac{\vec{n}\vec{r}}{r^{2}-\left( \vec{n}\vec{r}%
\right) ^{2}}\,,  \notag \\
&&\vec{n}=\left( \sin \theta _{k}\cos \alpha ,\sin \theta _{k}\sin \alpha
,\cos \theta _{k}\right)  \label{61}
\end{eqnarray}%
(the string of the potential is directed along the vector $\vec{n}$) and the
wave function is given by the following expression:

\begin{eqnarray}
&&\psi _{k}\left( \vec{r}\right) =e^{i\mu _{1}\Omega _{1}-i\mu _{1}\alpha
}\psi _{k}^{\left( 0\right) }\left( \vec{r}\right) \,,  \label{62} \\
&&\psi _{k}^{\left( 0\right) }\left( \vec{r}\right) =e^{-i\mu _{1}\pi +i\mu
_{1}\varphi }\underset{L}{\sum }\left( 2L+1\right) i^{L}e^{i\delta
_{L}}P_{L}^{-\mu _{1},\mu _{1}}\left( \theta _{kr}\right) R_{k,L}\left(
r\right) \,,  \label{63} \\
&&\tan \Omega =\frac{\sin \theta _{k}\sin \left( \varphi -\alpha \right) }{%
\sin \theta _{k}\cos \theta \sin \left( \varphi -\alpha \right) -\sin \theta
\cos \theta _{k}}\,.  \label{64}
\end{eqnarray}

Thus, the wave function (\ref{62}) is a solution of the Schr\"{o}dinger
equation for the scattering of an electron by potential (\ref{61}) with the
initial momenta of the electron directed along the $z$ axis. In addition,
the wave function (\ref{63}) is a solution of the same problem in the case
of a string directed along the $z$ axis. The scattering amplitude has the
form ($\theta \neq 0$,$\,\pi $)%
\begin{eqnarray}
&&f\left( \theta ,\varphi \right) =e^{i\mu _{1}\Omega _{1}-i\mu _{1}\alpha
}f^{\left( 0\right) }\left( \theta ,\varphi \right) \,,  \label{65} \\
&&f^{\left( 0\right) }\left( \theta ,\varphi \right) =\frac{e^{-i\mu _{1}\pi
+i\mu _{1}\varphi }}{2ik}\underset{L}{\sum }\left( 2L+1\right) P_{L}^{-\mu
_{1},\mu _{1}}\left( \theta \right) e^{2i\delta _{L}}\,,  \label{66}
\end{eqnarray}%
where $f^{\left( 0\right) }$ is the amplitude of electron scattering by the
potential with a string directed along the $z$ axis.\footnote{%
The expression (\ref{65}) for the scattering amplitude was obtained by
Zwanziger \cite{2} with the help of the group-theory method.}

Therefore, as has been said in Introduction, the dependence of the wave
function and scattering amplitude on the direction of a string (on $\vec{n}$%
) enters only a non-essential phase multiplier. The cross-section of
scattering, however, does not depend on the direction of a string.

A solution of the (seeming) paradox indicated in Introduction and related
with the appearance of the phase multiplier $e^{-i\mu _{1}\left( \Omega
-\alpha \right) }$ consists in the equality%
\begin{equation}
\vec{A}^{\left( S,n\right) }-\vec{A}^{\left( S\right) }=ie^{-i\mu _{1}\left(
\Omega -\alpha \right) }\vec{\partial}e^{i\mu _{1}\left( \Omega -\alpha
\right) }\,,  \label{67}
\end{equation}%
which is valid only for integer $\mu _{1}$. Given this, the relation $\vec{A}%
^{\left( S,n\right) }-\vec{A}^{\left( S\right) }=-\mu _{1}\vec{\partial}%
\Omega $ neither holds true nor follows from (\ref{67}), since $\Omega $ is
a discontinuous function. Representation (\ref{67}) is nevertheless
sufficient for us to transform the difference of potentials $\vec{A}^{\left(
S,n\right) }-\vec{A}^{\left( S\right) }$ out of the Schr\"{o}dinger equation
with the help of a gauge (however, not a gradient one) phase transformation
of the form (\ref{22}).

Let us briefly examine the problem of electron scattering by the Dirac
potential%
\begin{equation}
A_{k}^{\left( D\right) }\left( \vec{r}\right) =-\mu \frac{\varepsilon
_{3ki}r_{i}}{r\left( r-z\right) }\,.  \label{68}
\end{equation}

One can easily see that the expression $J^{2}$ takes the form%
\begin{equation}
J^{2}=-\frac{1}{\sin \theta }\partial _{\theta }\left( \sin \theta \partial
_{\theta }\right) +\frac{1}{\sin ^{2}\theta }\left[ \left( -i\partial
_{\varphi }+\mu \right) ^{2}+2\mu \left( -i\partial _{\varphi }+\mu \right)
\cos \theta +\mu ^{2}\right] \,,  \label{69}
\end{equation}%
and that its eigenfunctions are given by expressions (\ref{44}), with the
following parameters $\alpha $, $\beta $:%
\begin{equation}
\alpha =\left| -m+2\mu \right| \,,\;\beta =\left| m\right| \,.  \label{70}
\end{equation}

Repeating literally the considerations presented in the case of the
potential $\vec{A}^{\left( S\right) }$, we find that there exists a unique
self-adjoint operator $J^{2}$ with a ``non-singular'' domain corresponding
to the differential expression (\ref{69}). If $2\mu $ is non-integer, then
the wave function turns to zero on the string, which implies that the
electron ``senses'' the string.

Let us further examine the case of integer $2\mu $. A complete system of
eigenfunctions $U_{m,l}\left( \theta ,\varphi \right) $ of the operator $%
J^{2}$ is given by expression (\ref{44prime}) with parameters (\ref{70}), $%
l=0,1,2\ldots \, $. We will present them in the form%
\begin{eqnarray}
&&U_{m,l}\left( \theta ,\varphi \right) =e^{-im\varphi }P_{L}^{m^{\prime
},\mu }\left( \theta \right) =e^{-i\mu \varphi }T_{L}^{m^{\prime },\mu
}\left( \varphi ,\theta ,0\right) \,,  \label{71} \\
&&m^{\prime }=m-\mu \,,\;L=l+\frac{\left| m\right| }{2}+\frac{\left| -m+2\mu
\right| }{2}\geq \left| \mu \right| \,.  \label{71prime}
\end{eqnarray}

By expanding the asymptotics of a plain wave in terms of an arbitrary
complete system of functions on a sphere, we find that the wave function of
electron scattering by the Dirac potential (\ref{69}) is given by%
\begin{equation}
\psi _{k}\left( \vec{r}\right) =\underset{l,m}{\sum }\left( -i\right)
^{L}\left( 2L+1\right) e^{i\delta _{L}+im\left( \varphi _{k}+\pi -\varphi
\right) }P_{L}^{m^{\prime },\mu }\left( \pi -\theta _{k}\right)
P_{L}^{m^{\prime },\mu }\left( \theta \right) R_{L,k}\left( r\right) \,.
\label{72}
\end{equation}

Using the addition rule for spherical functions, we present the expression
for the wave function in the coordinate system related to the old coordinate
system by Euler's angles $\left( \frac{\pi }{2}-\varphi _{k},\theta
_{k},\alpha +\frac{\pi }{2}\right) $:%
\begin{eqnarray}
&&\psi _{k}\left( \vec{r}\right) =e^{i\mu \Omega ^{\prime }}\underset{L}{%
\sum }i^{L}\left( 2L+1\right) P_{L}^{-\mu ,\mu }\left( \theta \right)
e^{i\delta _{L}}R_{L,k}\left( r\right) \,,  \label{73} \\
&&L=l+\left| \mu \right| \,,\;l=0,1,2,\ldots \,,  \notag \\
&&\tan \frac{\Omega ^{\prime }}{2}=\frac{2\sin \frac{\theta _{k}}{2}\cos
\frac{\theta }{2}\sin \left( \varphi -\alpha \right) }{\sin \frac{\theta _{k}%
}{2}\cos \frac{\theta }{2}\cos \left( \varphi -\alpha \right) -\cos \frac{%
\theta _{k}}{2}\sin \frac{\theta }{2}}\,.  \label{74}
\end{eqnarray}

The corresponding amplitude $\phi $ equals to ($\theta \neq 0,\pi $)%
\begin{equation}
\phi \left( \theta ,\varphi \right) =\frac{e^{i\mu \Omega ^{\prime }}}{2ik}%
\underset{L}{\sum }\left( 2L+1\right) P_{L}^{m^{\prime },\mu }\left( \theta
\right) e^{2i\delta _{L}}\,.  \label{75}
\end{equation}

Expressions (\ref{73}) and (\ref{75}) present the wave function and
scattering amplitude in the case of the potential (the $z$ axis is directed
along $\vec{k}$)%
\begin{equation}
\vec{A}^{\left( D,n\right) }\left( \vec{r}\right) =\mu \frac{\vec{n}\times
\vec{r}}{r\left( r-\vec{n}\vec{r}\right) }  \label{76}
\end{equation}%
and differ from the corresponding expressions in the case of potential (\ref%
{68}) only by the phase multiplier $\exp \left( i\Omega ^{\prime }\right) $.
Therefore, in this problem the cross-section of scattering does not depend
on the direction of the string, either. The dependence on the string appears
only as a non-essential phase multiplier, which is related with the
following equality:%
\begin{equation}
\vec{A}^{\left( D,n\right) }\left( \vec{r}\right) -\vec{A}^{\left( D\right)
}\left( \vec{r}\right) =ie^{-i\mu \Omega ^{\prime }}\vec{\partial}e^{i\mu
\Omega ^{\prime }}  \label{77}
\end{equation}%
(valid only for integer $2\mu $).

Let us now compare solutions (\ref{63}) and (\ref{73}) in the case $\theta
_{k}=0$, $\alpha =0$ . It is clear that in the case of integer $\mu =\mu
_{1} $ these solutions differ only by the phase multiplier $\exp \left( i\mu
_{1}\varphi \right) $. At the same time, it is precisely this multiplier
that connects the potentials of Dirac and Schwinger:%
\begin{equation}
\vec{A}^{\left( S\right) }\left( \vec{r}\right) -\vec{A}^{\left( D\right)
}\left( \vec{r}\right) =-\mu \frac{\vec{n}_{z}\times \vec{r}}{r^{2}-z^{2}}%
=ie^{-i\mu \varphi }\vec{\partial}e^{i\mu \varphi }\,.  \label{78}
\end{equation}

We arrive at the following conclusion: if the field of a monopole is
described by several strings, then the magnetic charge (flow) of each string
must be half-integer (for instance, it is only under this condition that an
equality of the kind (\ref{77}) holds true, which ensures the
non-observability of a string). In the case of Schwinger's potential, we
deal with two strings of equal charges, which implies the quantization
condition $\mu =n$. In the case of Dirac's potential, there is only one
string, and thus the quantization condition is $\mu =\frac{n}{2}$. If, for
some reasons, we wish to describe the field of a monopole using $k$ strings
with equal charges, the quantization condition will then be $\mu =k\cdot
\frac{n}{2}$.

\section{Conclusion}

Let us briefly state the results once again. In the case of electron
scattering by the potential of infinitely thin infinite and semi-infinite
solenoids, is has been demonstrated that in both of these problems there
exist unique self-adjoint operators with ``non-singular'' domains, which, in
view of physical reasons, must be identified with the Hamiltonians of the
corresponding problems. In addition, if the magnetic charges of the strings
(i.e., magnetic flows along the strings) do not conform to the quantization
rules then the electron experiences scattering by the strings.

If, however, the magnetic charges actually conform to the quantization rules
then the electron does not feel the presence of such a string. This result
is also explained by the fact that the potential of a thin solenoid and the
difference of the potentials of two half-solenoids can be presented in the
form $U^{-1}\partial U$ and thus transformed out of the Schr\"{o}dinger
equation with the help of a gauge (however, not a gradient one)
transformation of the form (\ref{22}).

\appendix

\section{Appendix}

In \cite{7}, the problem of electron scattering by a thin solenoid is solved
in the framework of a perturbation theory in $\Delta \mu $. The wave
function is presented in the form ($\mu =n-\Delta \mu $)%
\begin{equation}
\psi \left( \vec{\rho}\right) =\psi _{0}\left( \vec{\rho}\right) +\Delta \mu
\psi _{1}\left( \vec{\rho}\right) \,,\;\psi _{0}\left( \vec{\rho}\right)
=\left( -\right) ^{n}e^{-in\varphi }e^{-i\vec{k}\vec{\rho}}\,,  \label{A.1}
\end{equation}%
where $\psi _{1}$ is subject to the equation%
\begin{equation}
\left( -\Delta -k^{2}\right) \widetilde{\psi }_{1}=-\frac{2i}{\rho ^{2}}%
\partial _{\varphi }e^{-i\vec{k}\vec{\rho}}\,,\;\psi _{1}=\left( -\right)
^{n}e^{-in\varphi }\widetilde{\psi }_{1}  \label{A.2}
\end{equation}%
(the momentum $\vec{k}$ being directed along the $x$ axis). The
corresponding scattering amplitude is%
\begin{equation}
\left| f\right| ^{2}=\frac{\pi ^{2}\Delta \mu ^{2}}{\tan ^{2}\frac{\varphi }{%
2}}\,;  \label{A.3}
\end{equation}%
however, the correct result for a small $\Delta \mu $ is%
\begin{equation}
\left| f\right| ^{2}=\frac{\pi ^{2}\Delta \mu ^{2}}{\sin ^{2}\frac{\varphi }{%
2}}\,.  \label{A.4}
\end{equation}

The difference between (\ref{A.3}) and (\ref{A.4}) can be traced to the fact
that (\ref{A.3}) does not contain the contribution due to the partial wave $%
m=-n$, whereas this contribution is present in (\ref{A.4}). To explain the
absence of the wave $m=-n$ in (\ref{A.3}) (and in $\psi _{1}$), let us
represent the expansion $\psi _{1}$ in partial waves:%
\begin{equation}
\psi _{1}\left( \vec{\rho}\right) =\underset{m}{\sum }e^{im\varphi -in\frac{%
\pi }{2}}i^{m}F_{m}\left( \rho \right) \,.  \label{A.5}
\end{equation}%
Then, $F_{m}$ obeys the equation%
\begin{equation}
\left( \partial _{\rho }^{2}+\frac{1}{\rho }\partial _{\rho }-\frac{\left(
m+n\right) ^{2}}{\rho ^{2}}+k^{2}\right) F_{m}\left( \rho \right) =-\frac{%
2\left( m+n\right) }{\rho ^{2}}J_{m+n}\left( k\rho \right) \,.  \label{A.6}
\end{equation}

The\ solution of equation (\ref{A.6}) with the asymptotics of a dispersed
wave (regular at $\rho =0$) is given by\footnote{%
We have introduced the notation $J_{a;\mu }(x)=\left. \frac{\partial }{%
\partial \mu }J_{a+\mu }\left( x\right) \right| _{\mu =0}\,.$}%
\begin{equation}
F_{m}\left( \rho \right) =\left\{
\begin{array}{l}
i\frac{\pi }{2}J_{m+n}\left( k\rho \right) -J_{m+n;\mu }\left( k\rho \right)
\,,\;m+n>0\,, \\
-i\frac{\pi }{2}J_{m+n}\left( k\rho \right) +\left( -\right) ^{m+n}J_{\left|
m+n\right| ;\mu }\left( k\rho \right) \,,\;m+n<0\,, \\
0\,,\;m+n=0\,.%
\end{array}%
\right.  \label{A.7}
\end{equation}

It is easy to see that (\ref{A.5}) with the coefficient functions (\ref{A.7}%
) is identical with the expression%
\begin{equation}
\psi _{1}\left( \vec{\rho}\right) =\left( -\right) ^{n+1}e^{in\varphi }\cdot
2i\Delta \mu \int d\vec{\rho}^{\prime }G\left( \vec{\rho}-\vec{\rho}^{\prime
}\right) \frac{1}{\rho ^{\prime 2}}\partial _{\varphi ^{\prime }}e^{i\vec{k}%
\vec{\rho}^{\prime }}\,.  \label{A.8}
\end{equation}%
used in \cite{7}, where $G\left( \vec{\rho}\right) =\frac{i}{4}H_{0}^{\left(
1\right) }\left( k\rho \right) $.

Let us compare (\ref{A.7}) with the expressions for partial coefficients
that follow from the expansion in $\Delta \mu $ of the exact solution (\ref%
{18}). It is clear that in case $m+n\neq 0$ the expansion in $\Delta \mu $
of the exact solution leads precisely to expression (\ref{A.7}). However, in
case $m+n=0$ the exact solution leads to%
\begin{equation}
F_{-n}\left( \rho \right) =-i\frac{\pi }{2}H_{0}^{\left( 1\right) }\left(
k\rho \right) \,.  \label{A.9}
\end{equation}%
At the same time, (\ref{A.7}) implies $F_{-n}=0$.

The solution of this paradox is given by the fact that the exact equation
for $F_{m}$ (as well as for $\psi _{k}$) cannot be solved by a perturbation
theory in $\Delta \mu $.

Indeed, let us examine the equation for Bessel's functions $J_{\nu +\Delta
\mu }$:%
\begin{equation}
\left( \partial _{x}^{2}+\frac{1}{x}\partial _{x}-\frac{\left( \nu +\Delta
\mu \right) ^{2}}{x^{2}}+1\right) J_{\nu +\Delta \mu }\left( x\right) =0\,.
\label{A.10}
\end{equation}

If one solves this equation in the framework of a formal perturbation theory
in $\Delta \mu $, then the lowest correction in $\Delta \mu $ to $J_{0}$: $%
J_{\Delta \mu }\thickapprox J_{0}+F_{0}$ must obey the equation%
\begin{equation}
\left( \partial _{x}^{2}+\frac{1}{x}\partial _{x}+1\right) F_{0}\left(
x\right) =\frac{\Delta \mu ^{2}}{x^{2}}J_{0}\left( x\right) =0\,,
\label{A.11}
\end{equation}%
being a second-order equation in $\Delta \mu $. At the same time, the first
derivative $\left. \frac{\partial }{\partial \Delta \mu }J_{\Delta \mu
}\left( x\right) \right| _{\Delta \mu =0}$ does not vanish (and equals to $%
\frac{\pi }{2}N_{0}\left( x\right) $). This means that the r.h.s. of the
equation for $F_{0}$ must contain terms of first-order in $\Delta \mu $. To
construct a correct equation for $F_{0}$, one can proceed, for instance, in
the following way. Let us first present the equation for $F_{\varepsilon
}\,(J_{\varepsilon +\Delta \mu }\thickapprox J_{\varepsilon }+F_{\varepsilon
},F_{\varepsilon }\equiv \Delta \mu \frac{\partial }{\partial \varepsilon }%
J_{\varepsilon })$ and then proceed to the limit $\varepsilon \rightarrow 0$.

We obtain%
\begin{eqnarray}
&&\left( \partial _{x}^{2}+\frac{1}{x}\partial _{x}+1\right) F_{0}\left(
x\right) =\Delta \mu Q\left( x\right) =0\,,  \label{A.12} \\
&&Q\left( x\right) =\underset{\varepsilon \rightarrow 0}{\lim }\left[ \frac{%
2\varepsilon }{x^{2}}J_{\varepsilon }\left( x\right) -\frac{\varepsilon ^{2}%
}{x^{2}}\frac{\partial }{\partial \varepsilon }J_{\varepsilon }\left(
x\right) \right] \,.  \label{A.13}
\end{eqnarray}%
From a simple calculation, it follows that%
\begin{equation}
Q\left( x\right) =\underset{\varepsilon \rightarrow 0}{\lim }\frac{%
\varepsilon }{x^{2-\varepsilon }}=\frac{1}{x}\delta \left( x\right) \,,
\label{A.14'}
\end{equation}%
where the r.h.s. of (\ref{A.14'}) is understood as%
\begin{equation}
\int dxdyf\left( \vec{x}\right) Q\left( \rho \right) =2\pi f\left( 0\right)
\underset{0}{\overset{\infty }{\int }}\rho d\rho Q\left( \rho \right) =2\pi
f\left( 0\right) \,.  \label{A.15'}
\end{equation}

Therefore, the r.h.s. of \ (\ref{A.12}) does not vanish and the solution of
the corresponding equation is precisely given by $F_{0}=\Delta \mu \frac{\pi
}{2}N_{0}$.

The presented analysis shows that the correct equation for the coefficient
functions $F_{m}$ is%
\begin{equation}
\left( \partial _{\rho }^{2}+\frac{1}{\rho }\partial _{\rho }-\frac{\left(
m+n\right) ^{2}}{\rho ^{2}}+k^{2}\right) F_{m}\left( \rho \right) =-\frac{%
2\left( m+n\right) }{\rho ^{2}}J_{m+n}\left( k\rho \right) +\delta _{0,m+n}%
\frac{\delta \left( \rho \right) }{\rho }\,.  \label{A.16'}
\end{equation}

Accordingly, the correct equation for $\widetilde{\psi }_{1}$ has the form%
\begin{equation}
\left( -\Delta -k^{2}\right) \widetilde{\psi }_{1}\left( \rho \right) =-%
\frac{2i}{\rho ^{2}}\partial _{\varphi }e^{i\vec{k}\vec{r}}+\frac{1}{\rho }%
\delta \left( \rho \right) \,.  \label{A.17'}
\end{equation}

The additional (as compared to (\ref{A.2})) term in the r.h.s. of (\ref%
{A.17'}) does not appear in the formal perturbation theory in $\Delta \mu $;
however, it is precisely this term that leads to the appearance of the
missing partial wave with $m=-n$. One can easily see that $\psi _{1}$,
obtained as a solution of \ (\ref{A.17'}), is in agreement with the result
of expanding the exact solution (\ref{18}) in $\Delta \mu $.

\section{Appendix}

Let us now present a proof of the uniform boundedness of the quantity $a%
\widetilde{F}_{m,k}^{\prime }\left( a\right) /\widetilde{F}_{m,k}\left(
a\right) $ (formula (\ref{35})).

From (\ref{26}) and (\ref{31}), it follows that the function $\widetilde{F}%
_{m,k}$ obeys the equation%
\begin{eqnarray}
&&\left( \partial _{\rho }^{2}+\frac{2\left| m\right| +1}{\rho }\partial
_{\rho }+k^{2}-\frac{2m\mu }{\rho ^{2}}f_{a}\left( \rho \right) -\frac{\mu
^{2}}{\rho ^{2}}f_{a}^{2}\left( \rho \right) \right) \widetilde{F}%
_{m,k}(\rho )=0\,,  \label{A.14} \\
&&\widetilde{F}_{m,k}\left( 0\right) =1\,.  \notag
\end{eqnarray}

The regularizing function $f_{a}\left( \rho \right) $ will be subject to the
conditions%
\begin{eqnarray}
f_{a}\left( a\right) &=&1\,,\;f_{a}\left( 0\right) =0\,,\;\left| \varphi
_{a}\left( \rho \right) \right| \leq C\,,\;0\leq \rho \leq a\,,  \notag \\
\left| af_{a}^{\prime }\left( a\right) \right| &<&C_{1}\,,\;0\leq \rho \leq
a\,,\;\mathrm{for\;all\;}a\,,  \label{A.15}
\end{eqnarray}%
where the constants $C$, $C_{1}$ do not depend on $a$; besides,%
\begin{equation}
\varphi _{a}\left( \rho \right) \equiv \frac{a}{\rho }f_{a}\left( \rho
\right) \,.  \label{A.16}
\end{equation}

Conditions (\ref{A.15}) can be met, for instance, by the regularizing
function%
\begin{equation}
f_{a}\left( \rho \right) \equiv f\left( \frac{\rho }{a}\right) \,.
\label{A.17}
\end{equation}

It can be verified immediately that the differential equation and initial
condition (\ref{A.14}) can also be satisfied by a solution (provided that it
does exist) of the following integral equation:%
\begin{equation}
\widetilde{F}_{m,k}\left( x\right) =1+\frac{1}{a}\underset{0}{\overset{x}{%
\int }}dy\left( 1-\frac{y^{2\left| m\right| }}{x^{2\left| m\right| }}\right) %
\left[ \frac{m}{\left| m\right| }\cdot \mu \varphi _{a}\left( y\right) +%
\frac{\mu ^{2}}{2\left| m\right| }\frac{y}{a}\varphi _{a}^{2}\left( y\right)
-\frac{ak^{2}y}{2\left| m\right| }\right] \widetilde{F}_{m,k}\left( y\right)
\,.  \label{A.18}
\end{equation}

Let us solve (\ref{A.18}) by iterations:%
\begin{eqnarray}
&&\widetilde{F}_{m,k}\left( x\right) =1+\overset{\infty }{\underset{n=1}{%
\sum }}Y_{n}\left( x\right) \,,  \label{A.19} \\
&&\widetilde{F}_{m,k}\left( x\right) =\frac{1}{a}\underset{0}{\overset{x}{%
\int }}dy\left( 1-\frac{y^{2\left| m\right| }}{x^{2\left| m\right| }}\right) %
\left[ \frac{m}{\left| m\right| }\mu \varphi _{a}\left( y\right) +\frac{\mu
^{2}}{2\left| m\right| }\frac{y}{a}\varphi _{a}^{2}\left( y\right) -\frac{%
ak^{2}y}{2\left| m\right| }\right] Y_{n-1}\left( y\right) \,.  \label{A.20}
\end{eqnarray}

Since there exists the inequality%
\begin{equation}
\left| \frac{m}{\left| m\right| }\mu \varphi _{a}\left( y\right) +\frac{\mu
^{2}}{2\left| m\right| }\frac{y}{a}\varphi _{a}^{2}\left( y\right) -\frac{%
ak^{2}y}{2\left| m\right| }\right| <C_{2}\,,\;0\leq x\leq a\,,  \label{A.21}
\end{equation}%
where $C_{2}$ is a certain constant, independent of $m$ and $a$, we find
that $Y_{n}$ admits the following estimation:%
\begin{equation}
\left| Y_{n}\left( x\right) \right| \leq \frac{1}{n!}\left( C_{2}\frac{x}{a}%
\right) ^{n}.  \label{A.22}
\end{equation}

Therefore, the iteration series (\ref{A.19}) converges absolutely, whereas a
solution of equation (\ref{A.18}) does exist and coincide with the required
solution of the differential equation (\ref{A.19}) (because the second
solution of the differential equation (\ref{A.19}) is singular at zero). In
addition, $\widetilde{F}_{m,k}$ satisfies the condition (uniform in $m$ and $%
a$)%
\begin{equation}
\left| \widetilde{F}_{m,k}\left( x\right) \right| \leq \exp \left( C_{2}%
\frac{x}{a}\right) \,,\;0\leq x\leq a\,.  \label{A.23}
\end{equation}

Taking a first derivative of (\ref{A.19}), we obtain%
\begin{equation}
\widetilde{F}_{m,k}^{\prime }\left( x\right) =\frac{2\left| m\right| }{%
ax^{2\left| m\right| +1}}\underset{0}{\overset{x}{\int }}dyy^{2\left|
m\right| }\left( \frac{m}{\left| m\right| }\mu \varphi _{a}\left( y\right) +%
\frac{\mu ^{2}}{2\left| m\right| }\frac{y}{a}\varphi _{a}^{2}\left( y\right)
-\frac{ak^{2}y}{2\left| m\right| }\right) \widetilde{F}_{m,k}\left( y\right)
\,,  \label{A.24}
\end{equation}%
whence it follows that $a\widetilde{F}_{m,k}^{\prime }$ is also bounded
uniformly in $m$ and $a$:%
\begin{equation}
\left| a\widetilde{F}_{m,k}\left( x\right) \right| \leq C_{2}\exp \left(
C_{2}\frac{x}{a}\right) \,,\;0\leq x\leq a\,.  \label{A.25}
\end{equation}
\ Let us now introduce a function $\phi _{m,k}$,%
\begin{equation}
\widetilde{F}_{m,k}^{\prime }\left( x\right) =\exp \left\{ \frac{m}{\left|
m\right| }\cdot \frac{\mu }{a}\underset{0}{\overset{x}{\int }}dy\varphi
_{a}\left( y\right) \right\} \phi _{m,k}\left( y\right) \,.  \label{A.26}
\end{equation}%
The function $\phi _{m,k}$ obeys the differential equation%
\begin{equation}
\left[ \partial _{x}^{2}+\left( \frac{2m}{\left| m\right| }\frac{\mu }{x}%
f_{a}\left( x\right) +\frac{2\left| m\right| +1}{x}\right) \partial
_{x}+k^{2}+\frac{m}{\left| m\right| }\frac{\mu }{x}f_{a}^{\prime }\left(
x\right) \right] \phi _{m,k}\left( x\right) =0\,,\;\phi _{m,k}\left(
0\right) =1\,,  \label{A.27}
\end{equation}%
and the integral equation%
\begin{equation}
\phi _{m,k}\left( x\right) =1-\frac{1}{2ma}\underset{0}{\overset{x}{\int }}%
dy\left( 1-\frac{y^{2\left| m\right| }}{x^{2\left| m\right| }}\right) \left[
\left( \frac{m}{\left| m\right| }ayk^{2}+a\mu f_{a}^{\prime }\left( y\right)
\right) \phi _{m,k}\left( y\right) +\mu y\varphi _{a}\left( y\right) \phi
_{m,k}^{\prime }\left( y\right) \right] \,.  \label{A.28}
\end{equation}

From (\ref{A.23}) and (\ref{A.25}), it follows that $\left| \phi
_{m,k}\left( x\right) \right| $ and $\left| x\phi _{m,k}^{\prime }\left(
x\right) \right| $ are bounded by some constants independent of $m$ and $a$.
We can, therefore, check once again that solution (\ref{A.28}) does exist
and coincide with solution (\ref{A.27}). For instance, the term with $\phi
_{m,k}^{\prime }$ in (\ref{A.28}) can be associated with the inhomogenoeous
term of the equation (bounded by a constant independent of $m$ and $a$), and
we obtain, as a result, an equation of the kind (\ref{A.18}), which allows
one to carry out the same estimation.

Looking once again at (\ref{A.28}), we find that $\phi _{m,k}$ can be
presented in the form%
\begin{equation}
\phi _{m,k}=1+\frac{1}{m}\theta \left( x\right) \,,  \label{A.29}
\end{equation}%
where%
\begin{equation}
\left| \theta \left( x\right) \right| =\frac{1}{2a}\left| \underset{0}{%
\overset{x}{\int }}dy\left( 1-\frac{y^{2\left| m\right| }}{x^{2\left|
m\right| }}\right) \left[ \left( ayk^{2}\frac{m}{\left| m\right| }+a\mu
f_{a}^{\prime }\left( y\right) \right) \phi _{m,k}\left( y\right) +\mu
y\varphi _{a}\left( y\right) \phi _{m,k}^{\prime }\left( y\right) \right]
\right| \leq C_{3}\,,  \label{A.30}
\end{equation}%
with $C_{3}$ being independent of $m$ and $a$. Using (\ref{A.25}), (\ref%
{A.26}), (\ref{A.29}) and (\ref{A.30}), we finally conclude that the
quantity $a\widetilde{F}_{m,k}^{\prime }\left( a\right) /\widetilde{F}%
_{m,k}\left( a\right) $ is actually bounded uniformly in $m$ and $a$.

With the help of (\ref{A.24}), it is easy to make sure that $\widetilde{F}%
_{m,k}^{\prime }$ has the following asymptotics in $m$:%
\begin{equation}
\widetilde{F}_{m,k}\left( x\right) =\frac{\mu }{a}\cdot \frac{m}{\left|
m\right| }\varphi _{a}\left( x\right) e^{\frac{\mu }{a}\cdot \frac{m}{\left|
m\right| }\underset{0}{\overset{x}{\int }}dy\varphi _{a}\left( y\right)
}+O\left( \frac{1}{m}\right) \,,\;m\rightarrow \infty \,.  \label{A.31}
\end{equation}

Therefore, quantity (\ref{35}) admits the estimation%
\begin{equation}
U_{m,k}\left( a\right) \equiv a\frac{\widetilde{F}_{m,k}^{\prime }\left(
a\right) }{\widetilde{F}_{m,k}\left( a\right) }=\mu \cdot \frac{m}{\left|
m\right| }+O\left( \frac{1}{m}\right) \,.  \label{A.32}
\end{equation}

This estimation can also be obtained as follows. The function $U_{m,k}$
obyes the differential equation%
\begin{equation}
xU_{m,k}^{\prime }\left( x\right) +2\left| m\right| U_{m,k}\left( x\right)
+U_{m,k}^{2}\left( x\right) =2m\mu f_{a}\left( x\right) +\mu
^{2}f_{a}^{2}\left( x\right) -k^{2}x^{2}\,.  \label{A.33}
\end{equation}

If we suppose that $U_{m,k}\left( x\right) $ and $U_{m,k}^{\prime }\left(
x\right) $ are bounded as $m\rightarrow \infty $, then (\ref{A.33}) implies (%
\ref{A.32}).

All of the obtained results are confirmed by an explicit calculation for two
functions $f_{a}$, in case equation (\ref{A.14}) has a manifest solution:%
\begin{eqnarray*}
1)\; &&f_{a}=\frac{\rho }{a}\,, \\
&&\widetilde{F}_{m,k}\left( \rho \right) =e^{-\xi \sqrt{\mu ^{2}-a^{2}k^{2}}%
\frac{\rho }{a}}\Phi \left( \frac{1}{2}+\left| m\right| +\lambda \,,2\left|
m\right| +1\,;2\xi \sqrt{\mu ^{2}-a^{2}k^{2}}\frac{\rho }{a}\right) \,, \\
&&\lambda =\frac{\left| m\mu \right| }{\sqrt{\mu ^{2}-a^{2}k^{2}}}\,,\;\xi =%
\frac{\left| m\mu \right| }{m\mu }\,, \\
2)\; &&f_{a}=\frac{\rho ^{2}}{a^{2}}\,, \\
&&\widetilde{F}_{m,k}\left( \rho \right) =e^{-\frac{\left| \mu \right| }{2}%
\frac{\rho ^{2}}{a^{2}}}\Phi \left( \frac{1}{2}+\frac{\left| m\right| }{2}+%
\frac{m\mu }{2\left| \mu \right| }-\frac{a^{2}k^{2}}{4\left| \mu \right| }%
\,,\left| m\right| +1\,;\frac{\left| \mu \right| }{2}\frac{\rho ^{2}}{a^{2}}%
\right) \,,
\end{eqnarray*}%
where $\Phi \left( \alpha ,\beta ;x\right) $ is the degenerate
hypergeometric function \cite{9}.

\section{Appendix}

Let us now find the expansion for the asymptotics of a plain wave $e^{i\vec{k%
}\vec{r}}$. We examine the integral%
\begin{equation}
\int d\Omega e^{i\vec{k}\vec{r}}f\left( \theta ,\varphi \right)
\,,\;r\rightarrow \infty \,,  \label{A.34}
\end{equation}%
and calculate it by the method of a stationary phase. There are two
stationary points of the function $kr\left( \cos \theta \cos \theta
_{k}+\sin \theta \sin \theta _{k}\cos \left( \varphi -\varphi _{k}\right)
\right) $, namely,%
\begin{equation*}
\varphi =\varphi _{k}\,,\;\theta =\theta _{k}\,,\;\mathrm{and\;}\varphi
=\varphi _{k}+\pi \,,\;\theta =\pi -\theta _{k}\,.
\end{equation*}

It is easy to see that%
\begin{equation}
\int d\Omega e^{i\vec{k}\vec{r}}f\left( \theta ,\varphi \right) =\frac{2\pi
}{ikr}\left[ e^{ikr}f\left( \theta _{k},\varphi _{k}\right) -e^{-ikr}f\left(
\pi -\theta _{k},\varphi _{k}+\pi \right) \right] \,,\;r\rightarrow \infty
\,.  \label{A.35}
\end{equation}%
(\ref{A.35}) implies that the asymptotics of a plain wave can be presented
in the form%
\begin{equation}
e^{i\vec{k}\vec{r}}\underset{r\rightarrow \infty }{\rightarrow }\frac{2\pi }{%
ikr}\left[ e^{ikr}\delta \left( \Omega -\Omega _{\vec{k}}\right)
-e^{-ikr}\delta \left( \Omega -\Omega _{\vec{k}}\right) \right] \,,
\label{A.36}
\end{equation}%
where the $\delta $-function in (\ref{A.35}) is understood as the $\delta $%
-function on a sphere.

Let us now suppose that we have a complete, orthonormalized system of
functions $T_{i}\left( \theta ,\varphi \right) $ on a sphere. Then, the
asymptotics of a plain wave can be presented in the form%
\begin{equation}
e^{i\vec{k}\vec{r}}\underset{r\rightarrow \infty }{\rightarrow }\frac{2\pi }{%
ikr}\left[ e^{ikr}\underset{n}{\sum }T_{n}^{\ast }\left( \theta _{k},\varphi
_{k}\right) T_{n}\left( \theta ,\varphi \right) -e^{-ikr}\underset{n}{\sum }%
T_{n}^{\ast }\left( \pi -\theta _{k},\varphi _{k}+\pi \right) T_{n}\left(
\theta ,\varphi \right) \right] \,.  \label{A.37}
\end{equation}

In the case $T_{i}\left( \theta ,\varphi \right) =Y_{lm}\left( \theta
,\varphi \right) $, we have the well-known expansion for the asymptotics of
a plain wave in spherical functions.

Formula (\ref{A.37}) allows one to present expressions for the wave function
and scattering amplitude in the case of Dirac's and Schwinger's potentials
for arbitrary $\mu $ as a series of the form (\ref{52}), (\ref{57}) and (\ref%
{72}).

For the sake of completeness, let us also obtain the asymptotics of a plain
wave in the two-dimensional case. To this end, we apply the method of a
stationary phase to the integral%
\begin{equation}
\int d\varphi e^{i\vec{k}\vec{r}}f\left( \varphi \right) \,,\;\rho
\rightarrow \infty \,.  \label{A.38}
\end{equation}%
The function $k\rho \cos \left( \varphi -\varphi _{k}\right) $ has two
stationary points:\ $\varphi =\varphi _{k}$ and $\varphi =\varphi _{k}+\pi $%
; a simple integration yields%
\begin{equation}
\int d\varphi e^{i\vec{k}\vec{r}}f\left( \varphi \right) \underset{\rho
\rightarrow \infty }{\rightarrow }\sqrt{\frac{2\pi }{k\rho }}\left[
e^{ik\rho -i\frac{\pi }{4}}f\left( \varphi _{k}\right) +e^{-ik\rho +i\frac{%
\pi }{4}}f\left( \varphi _{k}+\pi \right) \right] \,.  \label{A.39}
\end{equation}

Therefore, a two-dimensional plain wave has the following asymptotics:%
\begin{equation}
e^{i\vec{k}\vec{r}}\underset{\rho \rightarrow \infty }{\rightarrow }\sqrt{%
\frac{2\pi }{k\rho }}\left[ e^{ik\rho -i\frac{\pi }{4}}\delta \left( \varphi
-\varphi _{k}\right) +e^{-ik\rho +i\frac{\pi }{4}}\delta \left( \varphi
-\varphi _{k}-\pi \right) \right] \,.  \label{A.40}
\end{equation}

If we have a complete orthonormalized system of functions $T_{n}\left(
\varphi \right) $ on the segment $\left[ 0,2\pi \right] $, then the
expansion of the asymptotics of a two-dimensional plain wave can be
presented in the form%
\begin{equation}
e^{i\vec{k}\vec{r}}\underset{r\rightarrow \infty }{\rightarrow }\sqrt{\frac{%
2\pi }{k\rho }}\left[ e^{ik\rho -i\frac{\pi }{4}}\underset{n}{\sum }%
T_{n}^{\ast }\left( \varphi _{k}\right) T_{n}\left( \varphi \right)
-e^{-ik\rho +i\frac{\pi }{4}}\underset{n}{\sum }T_{n}^{\ast }\left( \varphi
_{k}+\pi \right) T_{n}\left( \varphi \right) \right] \,.  \label{A.41}
\end{equation}

In the case $T_{n}\left( \varphi \right) =\frac{1}{\sqrt{2\pi }}e^{in\varphi
}$, we obtain Exp. (\ref{17}).

\end{document}